\documentclass[longauth]{aa}  
\usepackage[varg]{txfonts}

\usepackage{bm} 

\usepackage[T1]{fontenc}
\usepackage{ae,aecompl}
\usepackage{graphicx}   
\usepackage{amsmath}    
\usepackage{amssymb}    

\usepackage[x11names]{xcolor}

\usepackage{amsmath}	
\usepackage{mathtools}
\usepackage{amssymb}	
\usepackage{natbib}
\bibpunct{(}{)}{;}{a}{}{,} 

\usepackage{hyperref}

\hypersetup{
   colorlinks=true,
   linkcolor=blue,
   citecolor=blue,
   urlcolor=blue,
}

\begin{document}

\title{
\centering
~~~~~~~~~~The ALPINE-ALMA [CII] survey:
\newline Star-formation-driven outflows and circumgalactic enrichment in the early Universe}
\titlerunning{SF-driven outflows and CGM enrichment in the early Universe}

\author{M. Ginolfi\inst{\ref{inst1}}
        \and
        G. C. Jones\inst{\ref{inst2},} \inst{\ref{inst3}}
        \and
        M. B\'{e}thermin\inst{\ref{inst4}}
        \and
        Y. Fudamoto\inst{\ref{inst1}}
        \and
        F. Loiacono\inst{\ref{inst5}, } \inst{\ref{inst6}}
        \and
        S. Fujimoto \inst{\ref{inst7}, } \inst{\ref{inst8}}
        \and
        O. Le Fèvre \inst{\ref{inst4}}
                \and \newline
        A. Faisst \inst{\ref{inst9}}
                        \and
        D. Schaerer \inst{\ref{inst1}}
                        \and
        P. Cassata \inst{\ref{inst10}}
                                \and
        J. D. Silverman \inst{\ref{inst8}, } \inst{\ref{inst11}}
                                \and
Lin Yan  \inst{\ref{inst12}}
                                \and
P. Capak \inst{\ref{inst9}}
                                \and
S. Bardelli \inst{\ref{inst5}}
                                \and
M. Boquien \inst{\ref{inst14}}
                                \and
R. Carraro \inst{\ref{inst15}}
\and
M. Dessauges-Zavadsky \inst{\ref{inst1}}
\and
M. Giavalisco \inst{\ref{inst16}}
\and
C. Gruppioni \inst{\ref{inst5}}
\and
E. Ibar \inst{\ref{inst15}}
\and
Y. Khusanova \inst{\ref{inst4}}
\and \newline
 B. C. Lemaux \inst{\ref{inst17}}
        \and
 R. Maiolino \inst{\ref{inst2},} \inst{\ref{inst3}}
 \and
 D. Narayanan \inst{\ref{inst18}}
  \and
 P. Oesch \inst{\ref{inst1}}
                                \and
 F. Pozzi \inst{\ref{inst6}}
                                \and
G. Rodighiero \inst{\ref{inst10}}
                                \and
 M. Talia \inst{\ref{inst5}, } \inst{\ref{inst6}}
                                \and \newline
 S. Toft \inst{\ref{inst19},}  \inst{\ref{inst20}}
 \and           
  L. Vallini\inst{\ref{inst21}}
  \and
  D. Vergani \inst{\ref{inst5}}
    \and
  G. Zamorani \inst{\ref{inst5}}
}

\institute{
        Observatoire de Gen\`eve, Universit\`e de Gen\`eve, 51 Ch. des Maillettes, 1290 Versoix, Switzerland\\
        \email{michele.ginolfi@unige.ch}\label{inst1}
        \and
        Cavendish Laboratory, University of Cambridge, 19 J. J. Thomson Ave., Cambridge CB3 0HE, UK\label{inst2}
        \and
        Kavli Institute for Cosmology, University of Cambridge, Madingley Road, Cambridge CB3 0HA, UK\label{inst3}
        \and
        Aix Marseille Univ, CNRS, CNES, LAM, Marseille, France\label{inst4}
        \and
        Osservatorio di Astrofisica e Scienza dello Spazio - Istituto Nazionale di Astrofisica, via Gobetti 93/3, I-40129, Bologna, Italy\label{inst5}
        \and
        University of Bologna, Department of Physics and Astronomy (DIFA), Via Gobetti 93/2, I-40129, Bologna, Italy\label{inst6}
                \and
        Institute for Cosmic Ray Research, The University of Tokyo, Kashiwa-no-ha, Kashiwa 277-8582, Japan\label{inst7}
                        \and
        Department of Astronomy, School of Science, The University of Tokyo, 7-3-1 Hongo, Bunkyo, Tokyo 113-0033, Japan\label{inst8}
                                \and
        IPAC, California Institute of Technology, 1200 East California Boulevard, Pasadena, CA 91125, USA\label{inst9}
                                        \and
        University of Padova, Department of Physics and Astronomy Vicolo Osservatorio 3, 35122, Padova, Italy\label{inst10}
                                                \and
Kavli Institute for the Physics and Mathematics of the Universe, The University of Tokyo, Kashiwa, Japan 277-8583 (Kavli IPMU, WPI)\label{inst11}
        \and
Caltech Optical Observatories, Cahill Center for Astronomy and Astrophysics 1200 East California Boulevard, Pasadena, CA 91125, USA\label{inst12}
\and
Centro de Astronomia (CITEVA), Universidad de Antofagasta, Avenida Angamos 601, Antofagasta, Chile\label{inst14}
\and
Instituto de Fisica y Astronomia, Universidad de Valparaiso, Gran Bretana 1111, Playa Ancha, Valparaiso, Chile\label{inst15}
\and
Department of Physics and Astronomy, University of Massachusetts, Amherst, MA 01003, USA\label{inst16}
\and
Department of Physics, University of California, Davis, One Shields Ave., Davis, CA 95616, USA\label{inst17}
\and
Department of Astronomy, University of Florida, 211 Bryant Space Sciences Center, Gainesville, FL 32611 USA\label{inst18}
\and
Cosmic Dawn Center (DAWN) \label{inst19}
\and
Niels Bohr Institute, University of Copenhagen, Lyngbyvej 2, DK-2100 Copenhagen, Denmark \label{inst20}
\and
Leiden Observatory, Leiden University, PO Box 9500, 2300 RA Leiden, The Netherlands \label{inst21}
}

\date{Received XXX; accepted YYY}

\abstract{
        We study the efficiency of galactic feedback in the early Universe  
        by stacking the [C II] 158 $\mu$m emission in a large sample of normal star-forming galaxies at $4<z<6$ from the \textit{ALMA Large Program to INvestigate [C II] at Early times} (ALPINE) survey.
        Searching for typical signatures of outflows in the high-velocity tails of the stacked [C II] profile, 
        we observe (i) deviations from a single-component Gaussian model in the combined residuals and (ii) broad emission in the stacked [C II] spectrum, with velocities of $|v| \lesssim 500$  km s$^{-1}$.
        The significance of these features increases when stacking the subset of galaxies with star formation rates (SFRs) higher than the median (SFR$_{med} = 25$ ${\rm M_{\odot}}$ yr$^{-1}$), thus confirming their star-formation-driven nature.
        The estimated mass outflow rates are comparable to the SFRs, yielding mass-loading factors of the order of unity (similarly to local star-forming galaxies), 
        suggesting that star-formation-driven feedback may play a lesser role in quenching galaxies at $z>4$.
        From the stacking analysis of the datacubes, 
        we find that the combined [C II] core emission ($|v| < 200$  km s$^{-1}$) of the higher-SFR galaxies 
        is extended on physical sizes of $\sim$ 30 kpc (diameter scale), well beyond the analogous [C II] core emission of lower-SFR galaxies and the stacked far-infrared continuum.
        The detection of such extended metal-enriched gas, likely tracing circumgalactic gas enriched by past outflows, corroborates previous similar studies, confirming that baryon cycle and gas exchanges with the circumgalactic medium are at work in normal star-forming galaxies already at early epochs.
}


\maketitle

\section{Introduction}\label{sec:introduction}

Current models of galaxy formation widely agree on the key importance of stellar feedback in regulating the evolution of galaxies over cosmic time.
Massive stars ($\gtrsim8~{\rm M_{\odot}}$) emit copious high-energy photons during their lifetimes and inject energy and momentum in the surrounding gas through supernova (SN) explosions in the final stage of their evolution (see a review by \citealp{Woosley2002}).
These mechanisms can heat the gas and drive turbulent motions in the interstellar medium (ISM; e.g., \citealp{Dekel1986, MacLow1999, Hopkins2012, Hopkins2014}), reducing the star formation efficiency to the observed typical low values of a few percent of the free-fall time (e.g., \citealp{Kennicutt1998, Krumholz2005, Leroy2008, Leroy2013, Bigiel2011}).
Stellar feedback is also often invoked to explain the observed discrepancy between the measured galaxy luminosity (or stellar mass, $M_{\star}$) function (LF) and the dark matter (DM) halo mass function predicted by the standard cosmological model (e.g., \citealp{Benson2003, Silk2012, Behroozi2013}).
While the sharp exponential cut-off at the luminous end of the LF is usually ascribed to feedback from accreting black holes (BHs) in active galactic nuclei (AGNs; see e.g., \citealp{Bower2006, Cattaneo2009, Fabian2012}), SN feedback is thought to be the dominant mechanism in shaping the flat slope at the low-mass end of the LF (e.g., \citealp{Dekel1986, Heckman1990, Hopkins2014}).
In particular, intense episodes of star formation induce powerful SN-driven winds, which can efficiently accelerate the gas to hundreds of kilometers per second (see e.g., \citealp{Heckman2017}) and eventually expel it from the disk, (i) suppressing the star formation rate (SFR; 
e.g., \citealp{Somerville2015, Hopkins2016, Hayward2017}), and (ii) enriching the circumgalactic and intergalactic  medium (CGM and IGM) with heavy elements (e.g., \citealp{Oppenheimer2006, Oppenheimer2010, Pallottini2014}).
Observational evidence of stellar feedback has increased over the years (see \citealp{Veilleux2005, Erb2015} for thorough reviews). 
A widely adopted method to trace the kinematics of cold and warm outflowing gas consists in measuring the blueshift of metal absorption resonant lines in the rest-frame ultraviolet (UV) and optical bands, with respect to the systemic redshift (usually measured through strong optical emission lines). 
This technique has been extensively employed to characterise star-formation-driven outflows in both 
local (e.g., \citealp{Arribas2014, Chisholm2015,Chisholm2016,Chisholm2017, Cicone2016}) 
and distant galaxies, up to $z\lesssim3-4$ (e.g., \citealp{Shapley2003, Steidel2004, Steidel2010, Rubin2010, Talia2012, Rubin2014, Heckman2015, Talia2017}).
%

%
At higher redshifts, approaching the epoch of reionization, detecting outflows through absorption-line spectroscopy becomes challenging, mainly because of 
(i) increasingly weaker metal absorption features, and 
(ii) large uncertainties on the systemic redshifts, which cannot be obtained from Ly$\alpha$, whose line profile is strongly affected by intergalactic absorption and radiative transfer effects.
A possible way to overcome such limitations comes from the growing number of recent Atacama Large Millimeter/submillimeter Array (ALMA) observations of bright far-infrared (FIR) lines, such as for example [C II] 158 $\mu$m
(hereafter [C II])
and [O III] 88 $\mu$m at $z>4$ (see e.g., \citealp{Wagg2012, Capak2015, Maiolino2015, Inoue2016, Bradac2017, Hashimoto2018, Carniani2018, Matthee2019}). 
For instance, combining the redshift determined from [C II] with deep observed-frame optical spectra taken at DEIMOS/Keck, \cite{Sugahara2019}  constructed a high-signal-to-noise-ratio (high-S/N) composite far-UV spectrum of seven Lyman break galaxies (LBGs) at $z = 5-6$ (\citealp{Riechers2014, Capak2015}).
\cite{Sugahara2019} find  central outflow velocities  (i.e., values measured at  the line center corresponding to the bulk motion of the gas) of $v_{\rm out} \gtrsim$ 400 km s$^{-1}$ and maximum outflow velocities of about 800 km s$^{-1}$, 
highlighting an increase (by a factor $> 3$) with respect to galaxies at lower redshifts (see \citealp{Sugahara2017}).
%

%
To probe star-formation-driven outflows in the early Universe, an alternative method to rest-frame FUV absorption line spectroscopy consists in studying the broad wings in the high-velocity tails of FIR-line spectra, similarly to what is commonly done for luminous AGN-driven outflows (see e.g., \citealp{Maiolino2012, Feruglio2018, Decarli2018, Bischetti2018, Stanley2019}).
Unfortunately, even significant investments of ALMA time ($\lesssim$ 1 hour; see e.g., \citealp{Capak2015}) do not provide sufficiently good spectra to analyze in detail the weak broad components
of FIR lines in individual "normal"%
\footnote{
        Following the commonly adopted nomenclature in the literature, we use \textit{normal} when referring to galaxies on the \textit{star-forming Main Sequence} (e.g., \citealp{Brinchmann2004, Noeske2007, Daddi2010, Rodighiero2011, Speagle2014}).
} 
star-forming galaxies at $z>4$, and stacking of large samples would be needed.
Some indications of the discovery potential of the FIR-line stacking analysis come from recent results by \cite{Gallerani2018}, 
who found flux excesses at  about $v\pm 500$ km s$^{-1}$ in the stacked residual [C II]-spectrum of a small sample of nine galaxies at $z\sim5-6$ (\citealp{Capak2015}), likely ascribed to broad wings tracing star-formation-driven outflows.
\newline
\newline
Aiming to improve our understanding of galactic feedback at early epochs, we explore the efficiency of star-formation-driven outflows through a stacking analysis of [C II] emission lines in a large sample of normal galaxies at $4 < z < 6$, from our \textit{ALMA Large Program to Investigate C$^+$ at Early Times} (ALPINE) survey (\citealp{LeFevre2019a}; B\'{e}thermin et al. 2019; Faisst et al. 2019; see a short description of the survey in Sect. \ref{sec:Observations}). 
On the one hand, the diversity of ALPINE galaxies (almost 2 dex in SFR and $M_{\star}$ are spanned across the main sequence) and the wealth of ancillary multi-wavelength photometric data (from UV to FIR) enable us to investigate primary dependencies of stellar outflows on galaxy physical properties.
On the other hand, the large statistics provided (the number of [C II]-detected galaxies used for the stacking is approximately six-fold higher than similar previous  studies; see \citealp{Gallerani2018}) yields enough sensitivity to (i) map the spatial extension of the outflowing gas and (ii) constrain the circumgalactic enrichment on scales of a few tens of kiloparsec, providing new critical pieces of information on the baryon cycling physics that drives the evolution of high-$z$ galaxies.
\newline
\newline
The paper is organized as follows. 
In Sect. \ref{sec:Observations}
we describe the ALPINE survey and the data-reduction process, 
while in Sect. \ref{sec:Results} we describe the methods of our analysis and report the results.
Section \ref{sec:discussion} contains a discussion on the implications of our findings, and Conclusions are summarized in Sect. \ref{sec:conclusions}.
\newline
Throughout the paper, we assume a flat Universe with $\rm \Omega_{m} = 0.3$, $\rm \Omega_{\Lambda} = 0.7$ and $H_0 = 70$ km s$^{-1}$ Mpc$^{-1}$, and adopt a \textit{Chabrier} initial mass function (IMF; \citealp{Chabrier2003}).

\section{Sample and observations}\label{sec:Observations}

\begin{figure*}
        \centering
        \includegraphics[width=1\textwidth]{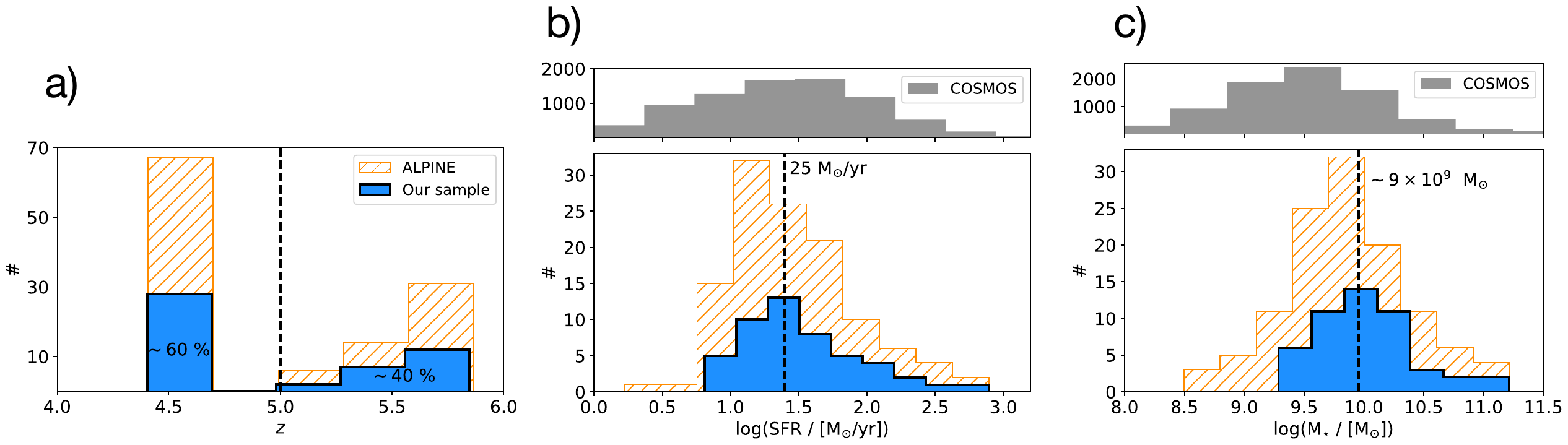}
        \caption{
                Redshift (a), SFR (b), and $M_{\star}$ (c) distributions of galaxies in the ALPINE survey (orange) and galaxies in our final sample (blue), drawn from ALPINE and used in this work (see Sect. \ref{sec:contaminants}).
                For comparison, in panels (b) and (c) we show the SFR and $M_{\star}$ distributions of all COSMOS galaxies with photometric redshift in the ALPINE range $z = [4.4 - 5.8]$ (from \citealp{Laigle2016}).
                The gap in the redshift distribution is due to the original ALPINE sample selection, tailored to avoid a prominent atmospheric absorption at $\sim 325$ GHz in ALMA band 7. 
                The black dashed lines in panels (b) and (c) represent the median SFR and $M_{\rm star}$ of galaxies in our ALPINE-drawn sample.   
        }
        \label{fig:physical_properties}
\end{figure*}

ALPINE is an ALMA large program (PI: O. Le Fèvre; \citealp{LeFevre2019a}; B\'{e}thermin et al. 2019; Faisst et al. 2019) designed to measure [C II] and the FIR-continuum emission for a representative sample of 118 normal galaxies at $z = [4.4 - 5.8]$. 
This enables extensive studies of the ISM and dust properties, as well as kinematics, and dust-obscured star formation in a representative population of high-$z$ galaxies, with template-fitting derived SFR $\gtrsim 5 ~ {\rm M_{\odot}}~ {\rm yr}^{-1}$ and stellar masses in the range $M_{\star} \sim 10^{8.5} - 10^{11} ~ {\rm M_{\odot}}$ (see Fig. \ref{fig:physical_properties}).
All galaxies
have reliable  optical spectroscopic redshifts coming from extensive campaigns at the Very Large Telescope (VUDS: \citealp{LeFevre2015, Tasca2017}) and Keck (Keck-COSMOS: \citealp{Hasinger2018}, Capak et al. in prep.),
and benefit from a wealth of ancillary multi-wavelength photometric data (from UV to FIR; see Faisst et al. in prep.).  
This makes ALPINE one of the currently largest panchromatic samples to study the physical properties of normal high-$z$ galaxies (see a discussion in \citealp{LeFevre2019a}, and  \citealp{Faisst2019}).
\newline
\newline
The overall ALMA observational strategy/setup and details on the data reduction steps (including data quality assessment) are comprehensively discussed in B\'{e}thermin et al. in prep.; however, a short summary of relevant information is reported here.
ALMA observations were carried out in Band 7 during Cycles 5 and 6, and completed in February 2019.
Each target was observed for between 20 minutes and 1 hour of exposure time, 
with phases centered at the rest-frame UV positions of the sources.
One spectral window was centered on the [C II] expected frequencies, according to the spectroscopic redshifts extracted from UV spectra, while other side bands were used for continuum measurements. 
The data were calibrated using the Common Astronomy Software Applications package (\texttt{CASA}; \citealp{McMullin2007}), version 5.4.0, and additional flagging of bad antennae was performed in a few cases (see B\'{e}thermin et al. in prep.).
Continuum maps were obtained running the \texttt{CASA} task \texttt{clean} (multi-frequency synthesis mode) over the line-free visibilities in all spectral windows, while [C II] datacubes were generated from the continuum-subtracted visibilities, with 500 iterations and a S/N threshold of $\sigma_{\rm{clean}}=3$ in the \texttt{clean} algorithm.
%
We chose a natural weighting of the visibilities, a common pixel size of 0.15$''$ and a common spectral bin of 25 km s$^{-1}$ (which is the best compromise in terms of number of spectral elements to resolve the line and S/N per channel).
The median sensitivity (in the spectral regions close to [C II] frequencies) reached by the cubes in our sample is $\sim 0.35$ mJy/beam for a 25 km s$^{-1}$ spectral channel, while the overall distribution ranges between 0.2 and 0.55 mJy/beam per channel with the same velocity binning. 
Such a variation (notwithstanding similar integration times) is mainly driven by the redshift range covered by our targets and the evolving atmospheric transmissivity function in ALMA Band 7 (see B\'{e}thermin et al. in prep.).
%
%
%
The typical angular resolution of the final products, computed as the average circularized beam axis, is 0.9$''$ ($\sim 5.2 - 6$ kpc in the redshift range $z = 4.4 - 5.8$), with values ranging 0.8$''$-1$''$.
%
%
%
%
%
%
In B\'{e}thermin et al. in prep., we discuss in detail the methods adopted to extract continuum and $\rm [C~ II]$ information from ALPINE observations, 
while we refer to other forthcoming works for an overview of ALPINE-related results.
%
%
%
The data set consists of 75  robustly [C II]-detected galaxies%
with S/N $> 3.5$ (i.e., the threshold at which our simulations indicate a 95\% reliability; see B\'{e}thermin et al. in prep.) calculated as the ratio between peak fluxes and  rms in optimally extracted%
\footnote{Velocity-integrated [C II] maps were created in an iterative way, allowing for (i) slight (astrometry-corrected) spatial offsets (<1$''$) between the [C II] and rest-frame UV centroids, and (ii) spectral shifts between [C II] line and expected frequencies from UV spectra (see B\'{e}thermin et al. in prep., for details).} 
[C II] velocity-integrated maps. 
We note that, as discussed in Sect. \ref{sec:Results}, in this work we exclude from our analysis $\sim 30\%$ of the sample, consisting of merging systems.

\section{Analysis and Results}\label{sec:Results}

[C II] is the brightest line in the FIR spectra of star-forming galaxies and has been exploited to trace AGN-driven outflows revealed by the presence of broad wings in the spectra of luminous high-$z$ QSOs (see e.g., \citealp{Maiolino2012,Cicone2015, Janssen2016,Feruglio2018,Decarli2018, Bischetti2018, Stanley2019}).
In normal galaxies, where outflows are expected to be powered to a greater extent by stellar feedback than by AGN activity, broad wings are expected to be less prominent and weaker (see, e.g., a review by \citealp{Heckman2017});
therefore, even in the deepest currently available $\rm [C ~II]$ observations at $z\gtrsim4$, the sensitivity is generally not adequate to detect weak broad components in individual objects (e.g., \citealp{Capak2015, Maiolino2015, Gallerani2018, Fujimoto2019}).
%

%
To explore the efficiency of galactic feedback at play in normal star-forming galaxies in the early Universe, we performed a stacking analysis of the [C II] emission in a sample of galaxies (see Sect. \ref{sec:contaminants}) drawn from the ALPINE survey (see Sect. \ref{sec:Observations}).
The stacking technique enables us to substantially increase the sensitivity in the combined spectra and cubes. It therefore holds a significant discovery potential as shown in \cite{Bischetti2018} and \cite{Gallerani2018}, who successfully carried out the stacking of a QSO sample at $4.5 < z < 7$, and a small sample of normal galaxies at $z \sim 5$, respectively.
%
%
%
%
In the following we describe the methods used to extract, align, and stack [C II] spectra and cubes of our galaxies, and report the results.

\subsection{\bf Methods and stacking analysis}\label{sec:methods}

Our analysis is based on three different procedures (described in the following paragraphs), each of them providing complementary information:
\newline
1) Stacking of the residuals, computed by subtracting a single-component Gaussian fit to each [C II] spectrum (Sect. \ref{sec:residuals}). This procedure is needed to test whether or not a single-Gaussian component is sufficient to describe (on average) our [C II] spectra.
\newline
2) Stacking of the [C II] spectra (Sect. \ref{sec:spectra}) to verify the improvement gained in describing the combined spectrum with a two-component Gaussian model, and to compute the typical outflow properties (e.g., velocity and mass of the neutral atomic gas).
\newline
3) Stacking of the [C II] cubes (Sect. \ref{sec:cubes}), to obtain information on the typical  spatial distribution of the [C II] emission, both at low- and high-velocities.       

\subsubsection{Extraction of spectra and alignment}\label{sec:extraction} 

To extract the [C II] spectra of our galaxies, we used 2D apertures defined by the pixels contained within the $2\sigma$-levels of our optimally extracted [C II] velocity-integrated maps (see B\'{e}thermin et al. in prep.).
Rather than adopting a common fixed aperture, this has the advantage of taking into account variable morphologies and/or extensions of the gas, in order to include most of the flux coming from the total [C II]-emitting region, and to minimize the addition of noise. 
However, as discussed in Sect. \ref{sec:spectra}, we also tested fixed and smaller apertures.

Before stacking, we align the spectral axes of both spectra and cubes according to their [C II] observed frequencies: 
we set as a common "zero-velocity" reference the 25 km s$^{-1}$ -sized channel or slice centered (after interpolation) on the centroid frequency of the Gaussian fit.
%
The resulting distribution of the number of objects per spectral element (as shown in the top panels of  Figs. \ref{fig:residuals_1} and \ref{fig:residuals_2}) is not uniform along the full velocity range and declines starting from a few hundred  kilometres per second around the line, and is halved at about $\pm 1000$ km s$^{-1}$. 
These effects are mainly due to:        
(i) the exclusion of a few spectral channels flagged by the pipeline during the reduction steps, and more importantly 
(ii) spectral offsets between the {observed} [C II] redshifts and the {expected} redshifts as derived from rest-frame UV spectra originally used to center the spectral windows (see B\'{e}thermin et al. in prep. and Cassata et al., in prep., for technical details and a physical interpretation of the velocity offsets, respectively).
We also spatially align the [C II] cubes, centering them on the brightest pixel of [C II] velocity-integrated maps.
This procedure is preferred to choosing the phase center (coincident with the centroid of rest-frame UV positions of the sources) as a common spatial reference point, since a few sources show small spatial offsets ($< 1''$, whose physical interpretation will be discussed in another paper) between the [C II] and optical images centroids%
\footnote{For the sake of clarity we repeated our analysis leaving the phase centers as common spatial reference points, and the results of cube-stacking are identical within the errors. 
        The lack of evident deviation is due to the fact that only a small fraction ($<10\%$) of our sample is affected by small offsets (<1$''$) between [C II] and rest-frame UV (see a discussion in a similar analysis by \citealp{Fujimoto2019}).}.

\subsubsection{Exclusion of possible contaminants} 
\label{sec:contaminants}

As explained in Sect. \ref{sec:methods} and discussed in the following paragraphs, we are interested in revealing deviations from a single-component Gaussian model and flux excesses in the high-velocity tails of the stacked spectra and cubes possibly due to SF-driven winds.
Since these effects may be mimicked by companion galaxies and satellites in interacting systems (see discussions in e.g., \citealp{Gallerani2018, Fujimoto2019, Pallottini2019}), 
we excluded from our analysis 25 objects (corresponding to $\sim30\%$ of the [C II]-detected ALPINE sample) with signs of ongoing major or minor mergers;
for those systems a proper spatial or spectral deblending cannot be performed and any attempt does not guarantee the removal of possible contamination.
Such selection is based on a morpho-kinematic classification, described in detail in \cite{LeFevre2019a} and performed combining information from the ancillary multi-wavelength photometry and the ALMA products (e.g., velocity-integrated [C II] maps, 2D kinematics maps, and position-velocity diagrams; see Jones et al. in prep.).
We note that our exclusion of interacting systems does not prevent the sample from being somehow still contaminated by unresolved, HST/ALMA undetected,
faint satellites%
\footnote{We estimate a limit of $\lesssim 1.5$ ${\rm M_{\odot}}$ yr$^{-1}$ on their SFR, based on the  absolute UV magnitude limit of our sample.}.
As discussed in the following sections, some arguments suggest that this effect should not be significant, however a more robust solution to this issue is yet to be provided and may require future deeper and higher-resolution observations.
\newline
\newline
The final sample, drawn from ALPINE and used in this work, consists of 50 normal star-forming galaxies at redshift $4.4 < z < 5.8$, with SFR $\sim 5 - 600$ ${\rm M_{\odot}}$ yr$^{-1}$ and log($M_{\star}/{\rm M_{\odot}}$) $\sim 9-11$ (see Fig. \ref{fig:physical_properties}).
Stellar masses and star formation rates are derived  using a rich set of available ancillary data, including ground-based imaging observations from rest-frame UV to the optical, HST observations in the rest-frame UV, and Spitzer coverage above the Balmer break.
We used \cite{Bruzual2003} composite stellar population template fitting, using the \texttt{LePhare} code (\citealp{Arnouts1999, Ilbert2006}) with a large range in stellar ages, metallicities, and dust reddening.  
For further details we refer to Faisst et al. in prep., where a focused discussion on the ALPINE ancillary dataset and fitted physical properties (including characterization of the systematic uncertainties resulting from modeling assumptions) will be presented.

\subsection{\bf Combining the residuals}\label{sec:residuals}

Prior to searching for signatures of star-formation-driven outflows in the high-velocity tails of the stacked [C II] spectrum, we checked the {null hypothesis} that the [C II] line profiles of our galaxies are well (and completely) described by a single-Gaussian model.
We performed a simple standard procedure (see e.g., \citealp{Gallerani2018}) described as follows:
\begin{itemize}
        
        \item[(i)] we fit a single-Gaussian profile to each [C II] spectrum (where the peak flux, center velocity%
        \footnote{We note that since the spectra were spectrally centered and aligned at $z_{[C II]}$ (as discussed in Sect. \ref{sec:extraction}), the center velocity is by definition $0$ km s$^{-1}$.}
        , and full width at half maximum (FWHM) are free parameters), and compute the model value $G_i$, in each independent 25 km s$^{-1}$-sized spectral bin $i$;
        \item[(ii)] for each spectrum we compute the residuals $R_i$, by subtracting in each channel the best-fitting Gaussian model $G_i$ from the observed flux $F_i$, i.e., $R_i = F_i - G_i$;
        \item[(iii)] we combine the residuals performing a variance-weighted stacking:
        
        \begin{equation}\label{eq:stack_res}
                R^{\rm{stack}}_i = \dfrac{\sum_{k=1}^{N} R_{i, k} \cdot w_k}{\sum_{k=1}^{N} w_k} ,
        \end{equation}
        where $N$ is the number of galaxies contributing to each velocity bin, and the weighting factor $w_k$ is defined as $w_k=1/\sigma^2_k$, where $\sigma_k$ is the spectral noise associated with the spectrum $k$. 
        We compute $\sigma_k$ as the root mean square (rms) of the noise contained in each spectrum excluding channels in the velocity range $[-800: +800]$ km s$^{-1}$ around the center, 
        to avoid the [C II] emission line of the galaxies themselves.
        We found this range to be an optimal compromise between (i) having a large number of independent spectral bins to use for the determination of noise and (ii) conservatively excluding the velocity range usually found to be affected by stellar outflows. The effectiveness of this choice is probed 
         \textit{a posteriori} by our own results, since (as discussed in the following) no significant residuals are found at $|v| > 600-700$ km s$^{-1}$.
\end{itemize}

\begin{figure}
	\centering
	\includegraphics[width=0.9\columnwidth]{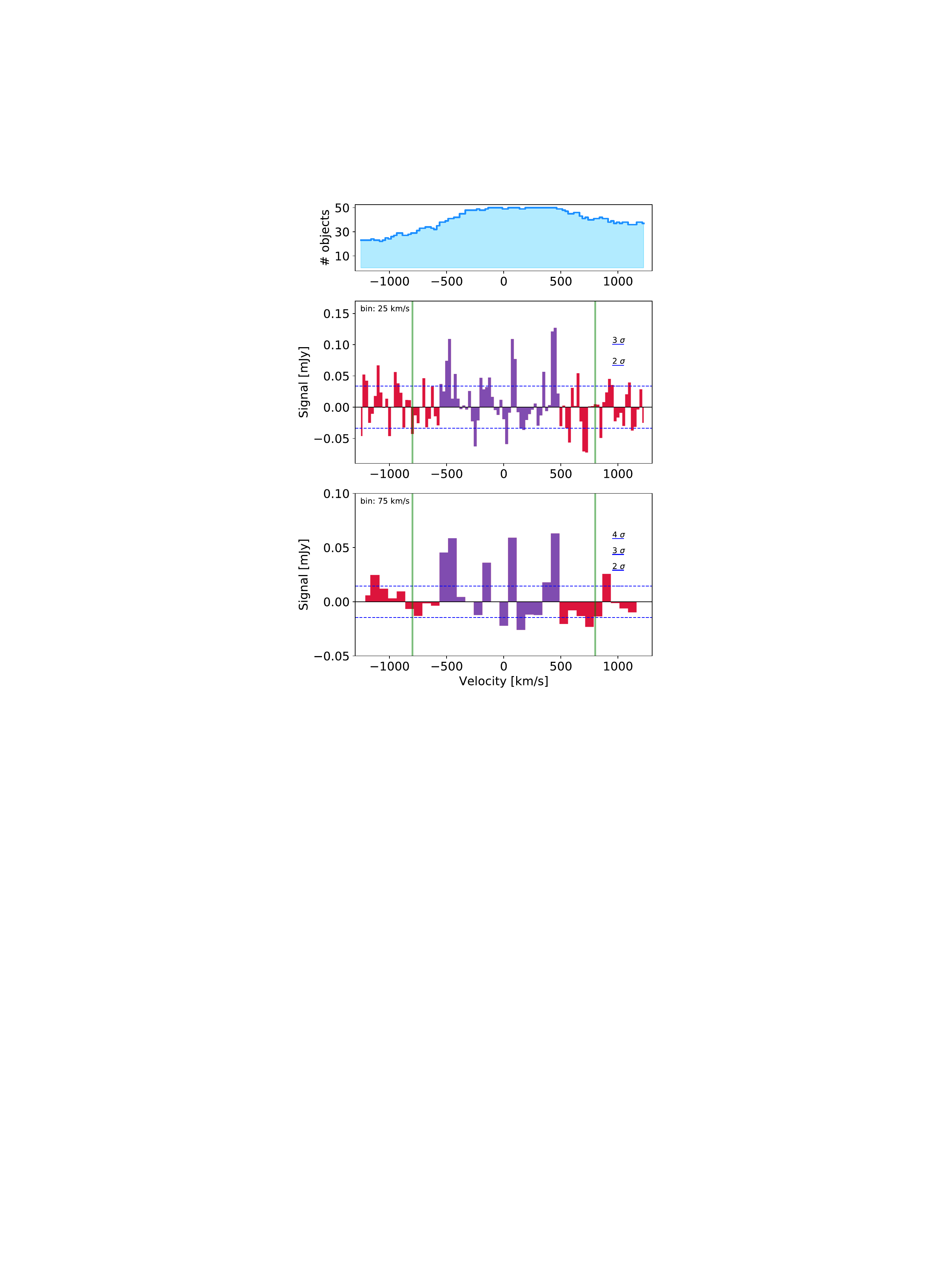}
	\caption{
		\textit{(Top)} Histogram containing the number of objects per channel contributing to the stacked flux.
		\textit{(Central lower panel)} Variance-weighted stacked [C II] residuals from a single-Gaussian fit in spectral bins of 25 km s$^{-1}$ (75 km s$^{-1}$). 
		The green solid lines at $\pm~800$ km s$^{-1}$ enclose the velocity range excluded for the estimation of spectral noise, while
		the blue dashed lines represent the spectral rms at $\pm 1\sigma$. 
		Channels in violet represent the channels in the velocity range enclosed by the outermost peaks at $\geq$ 3$\sigma$ in the 75 km s$^{-1}$-binned stacked residuals.
		This helps in visualizing the velocity interval affected by flux excesses. 
	}
	\label{fig:residuals_1}
\end{figure}

In Fig. \ref{fig:residuals_1} we show the resulting stacked residuals, $R^{\rm{stack}}_i$, where for each spectral channel $i$, we report in the top panel the number of sources contributing to the corresponding flux.
%
%
In the velocity range $v\sim[-500: +500]$ km s$^{-1}$ we find peaks of flux excess with significance  $> 3\sigma$ (where $\sigma$ is computed as the ratio between $R^{\rm{stack}}_i$ and $\sigma_k$) in single velocity bins 
(see violet bins, whose definition is reported in the caption of Fig. \ref{fig:residuals_1}), 
while the flux distribution in the stacked residuals at  larger velocities ($|v|>600$ km s$^{-1}$) is completely consistent with the noise. 
To facilitate the interpretation and improve the visualization, we re-bin the stacked residuals in channels of 75 km s$^{-1}$ (averaging over three contiguous spectral elements), revealing an increase of the flux excess significance up to $ 4 \sigma$ in the velocity range $v\sim[-500: +500]$ km s$^{-1}$.
We note that if our [C II] spectra were completely described by a single-Gaussian profile, the resulting flux from the stacked residuals should be simply consistent with random noise over the full velocity range.
\newline
\newline
To explore the origin of such observed deviations from a single-Gaussian profile and probe any connection with stellar feedback, we repeat the analysis described above dividing our sample into two SFR-defined bins, and analyzing each of them individually.
Specifically, we use the median SFR of galaxies in our sample (SFR$_{med} = 25$ ${\rm M_{\odot}}$ yr$^{-1}$; see Fig. \ref{fig:physical_properties}.b) as the threshold to create two equally populated 
subsamples of {low}-SFR galaxies (SFR $< 25$ ${\rm M_{\odot}}$ yr$^{-1}$) and {high}-SFR galaxies (SFR $\geq 25$ ${\rm M_{\odot}}$ yr$^{-1}$). %
%
%

\begin{figure*}
  	\centering
  	\includegraphics[width=0.9\textwidth]{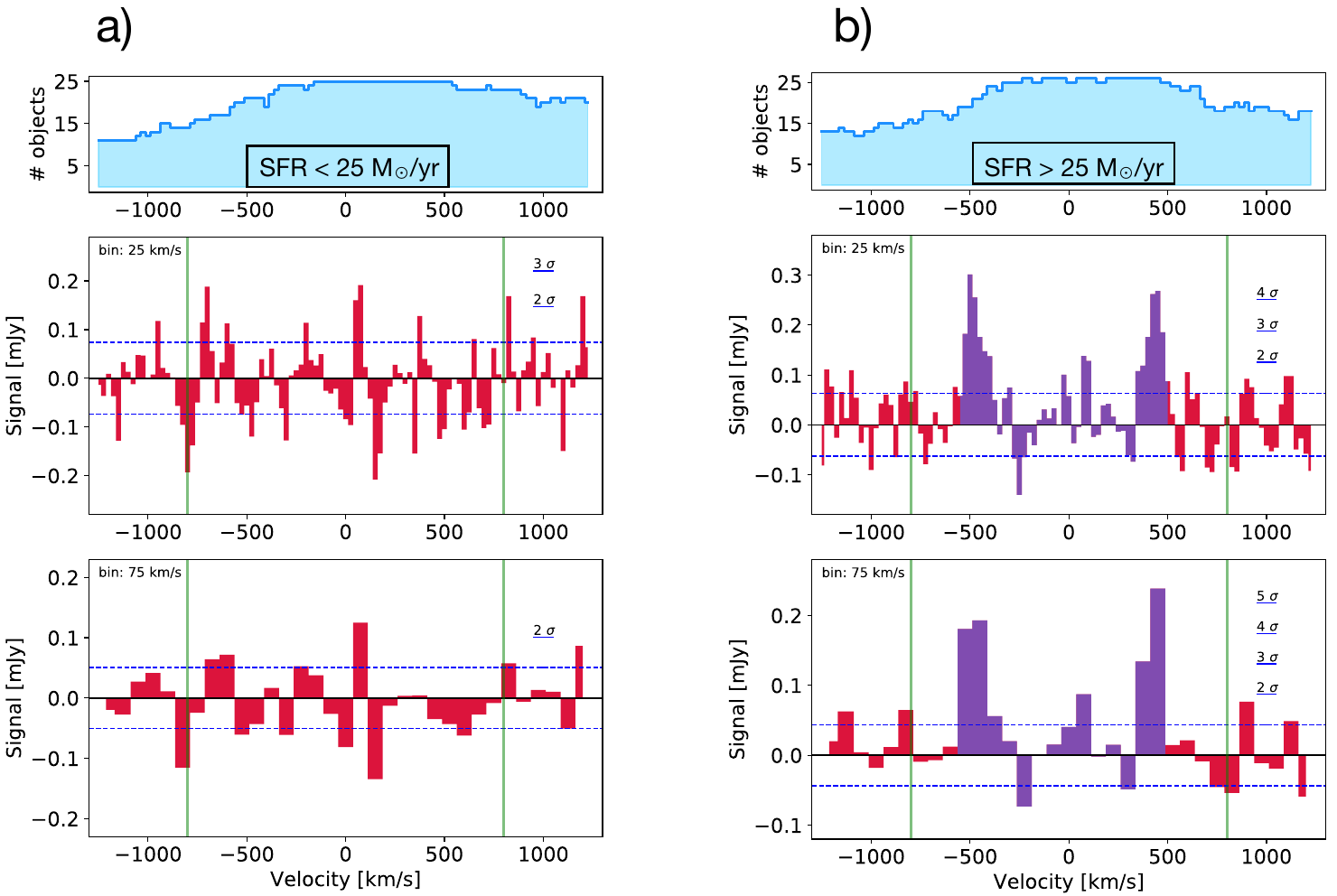}
  	\caption{Same description as in Fig. \ref{fig:residuals_1}. Here the stacked residuals are shown for the low-SFR (a), and the high-SFR (b) groups, respectively. 
  		While the stacked residuals are consistent with the noise in the low-SFR subsample, significant ($>4\sigma$) peaks of flux excess are detected for high-SFR galaxies.
  	}
  	\label{fig:residuals_2}
  \end{figure*}

We find that the stacked residuals of low-SFR galaxies do not show any clear sign of significant flux-excess over the entire velocity range, as shown in Fig. \ref{fig:residuals_2}.a. Channels at $v \sim [-500: +500]$ km s$^{-1}$ (where positive signal is detected when stacking the full sample; see Fig. \ref{fig:residuals_1}) are noise-dominated, with only few channels exceeding $2\sigma$; 

The flux excess at $|v| \lesssim  500$ km s$^{-1}$ in the stacked residuals of high-SFR galaxies is more distinct than in the stacked residuals of the full sample, 
        with (i) a larger number of connected velocity bins at S/N $>3\sigma$, and (ii) peaks reaching an increased significance of $4\sigma$ ($5\sigma$) in the 25 km s$^{-1}$ (75 km s$^{-1}$)-binned spectrum (see Fig. \ref{fig:residuals_2}.b).
        At lower velocities, $v\sim[-300: +300]$ km s$^{-1}$, the residuals appear flatter, 
        with some weak positive residuals around the zero and weak symmetric negative peaks at about $v \pm 250$ km s$^{-1}$ (see lower panel of Fig. \ref{fig:residuals_2}.b)%
        \footnote{
                These weak residuals are consist with the output of a single-Gaussian fit of a curve which is better described by the combination of a narrow and a broad Gaussian components (see Sect. \ref{sec:spectra}). 
                Indeed, in this case, a single-Gaussian fit would underestimate the center and overestimate the low-velocity flux in order to get some amplitude out into the high-velocity wings.}.
\newline
          
  \begin{figure}
  	\centering
  	\includegraphics[width=1\columnwidth]{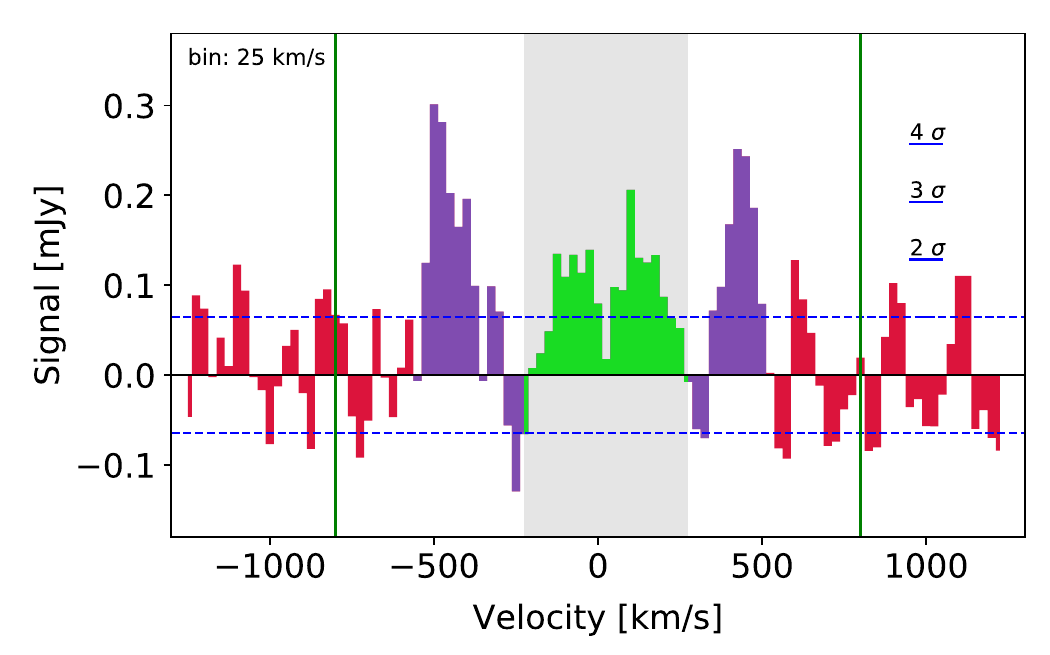}
  	\caption{
  		Similar to the central panel of Fig. \ref{fig:residuals_2}.b, the 25 km s$^{-1}$-binned stacked [C II] residuals of the high-SFR group are shown. However, here, for the five rotators in the subsample, $R_{i, k}$ (see Eq. \ref{eq:stack_res}) is calculated by subtracting a kinematic model to the observed spectra. We color in green the channels where the residuals left by the tilted-ring fit are non-null, specifically in the velocity range $v \sim [-225 : + 275]$  km s$^{-1}$, marked by the gray shaded region.
  	}
  	\label{fig:residuals_3}
  \end{figure}
  
        %
%
%

%

Therefore, the most star-forming galaxies in our sample contribute more to the deviation from a single-Gaussian profile, indicating a possible connection (on average) between the amount of SFR and  the observed deviation from a single-component Gaussian profile in the [C II] spectra of high-$z$ normal galaxies. 
Altogether, these findings suggest that star formation (or, more appropriately, star-formation-driven mechanisms) is likely to be responsible for producing the observed flux excess at the high-velocity tails of the stacked residuals.
\newline
\newline

\begin{figure*}
        \centering
        \includegraphics[width=1\textwidth]{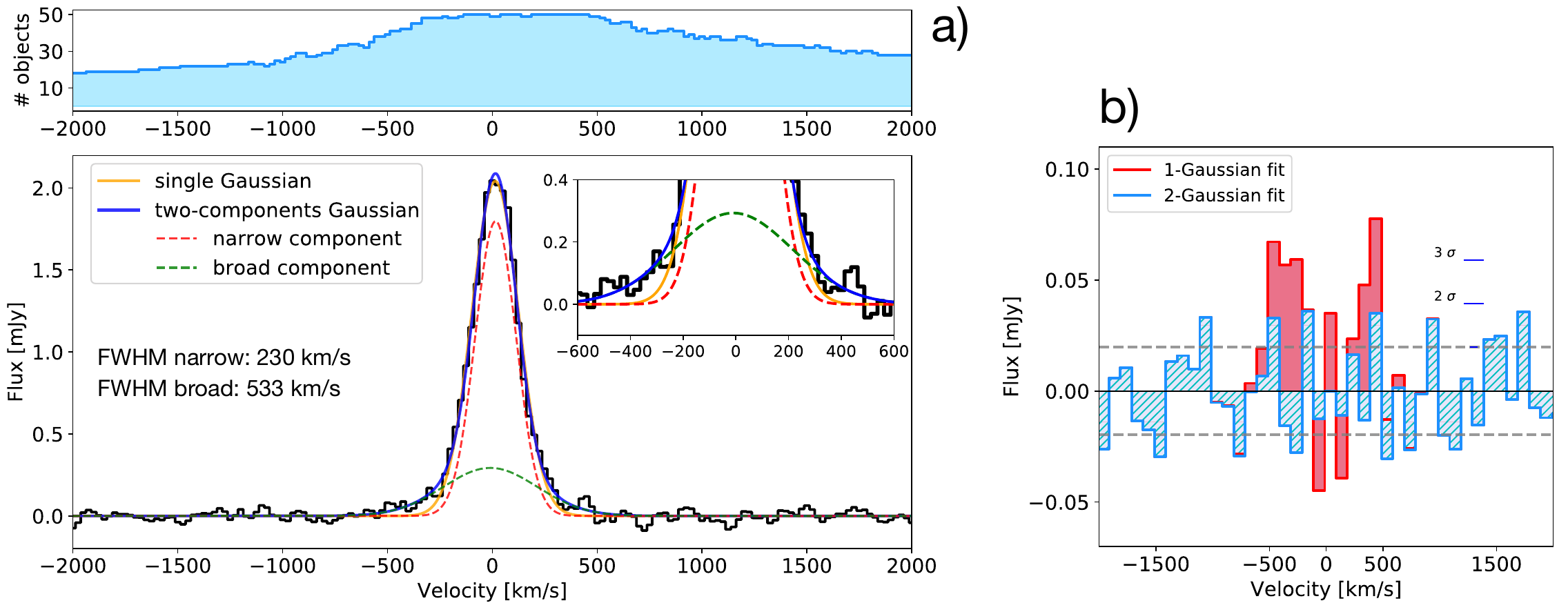}
        \caption{a) The variance-weighted stacked [C II] spectrum of all galaxies in our sample is shown, in velocity bins of 25 km s$^{-1}$. The orange (blue) line shows the single-Gaussian (two-Gaussian) best-fit. The red and the green line represent the narrow and broad components of the two-Gaussian model, respectively. A zoom of the velocity range $[-600 : +600]$ km s$^{-1}$ is shown in the inset. A histogram containing the number of objects per channel  contributing to the stacking is shown in the top panel.
                b) Residuals from the single-Gaussian (two-Gaussian) best-fit are shown in red (blue), in velocity bins of 50 km s$^{-1}$. 
        }
        \label{fig:spectrum_1}
\end{figure*}

\subsubsection*{Any contribution from rotating galaxies?}

While dispersion-dominated galaxies exhibit single-peak spectra, the double-horned profiles of rotating disks (see e.g., \citealp{Begeman1989, Daddi2010, deBlok2016, Kohandel2019}) are not well described by a single Gaussian. 
In addition, evidence for rotating disks has been found at high redshift (e.g., \citealp{DeBreuck2014, Jones2017, Talia2018, Smit2018}). 
Therefore, it is conceivable that the presence of rotating disks in our sample may contribute to the deviation from a single-Gaussian (see e.g., \citealp{Kohandel2019}) and to the production of the symmetric residuals seen in Fig. \ref{fig:residuals_2}.
However, we note that large rotational velocities of $|v|\sim500$ km s$^{-1}$ 
have been observed only in bright submillimeter galaxies (SMGs) and AGN-host galaxies, with intense SFRs $\gtrsim 1000 ~ {\rm M_{\odot}}$ yr$^{-1}$ and very broad FWHMs $\gtrsim 800$ km s$^{-1}$ (see e.g., \citealp{Carniani2013, Jones2017, Talia2018}),
and are unlikely to be produced by normal star-forming rotating galaxies (the median FWHM of [C II] profiles in our high-SFR galaxies is $\sim 250$  km s$^{-1}$).
\newline
\newline
To further explore this argument, we
use $^{\rm 3D}$\texttt{BAROLO} (a tool for fitting 3D tilted-ring models to emission-line datacubes that takes into account the effect of beam smearing; see \citealp{DiTeodoro2015})
to build kinematic models of five galaxies classified as rotators in the high-SFR group, for which we have enough independent spatial elements to obtain robust fits (Jones et al., in prep.).
The criteria adopted to classify rotators in ALPINE are discussed in detail in \cite{LeFevre2019a} (see also Sect. \ref{sec:contaminants}), and mainly require (i) smooth transitions between intensity channel maps, (ii) clear gradients in velocity field maps, (iii) tilted (straight) position--velocity diagrams projected along the major (minor) axis, (iv) possible double-horned profiles in the spectra and, (v) single components in ancillary photometric data.
We then repeat the residuals stack of our high-SFR galaxies, but now, for the five ALPINE rotators modeled with $^{\rm 3D}$\texttt{BAROLO}, we calculate 
$R_{i, k}$ (see Eq. \ref{eq:stack_res}) by subtracting the tilted-ring fit from the observed spectra rather than the Gaussian model.
The result is shown in Fig. \ref{fig:residuals_3}: 
while residuals resulting from the kinematic modeling (limited by our $\sim$1$''$ spatial resolution) are indeed visible (see green channels), 
%
these are more concentrated toward the common reference center, only affecting the velocity range $v \sim [-225 : + 275]$  km s$^{-1}$ (see gray shaded region).
%
This test suggests that the effect of unresolved kinematics in the spectra of our normal rotating galaxies should not have a significant impact on the residuals observed at $|v| \lesssim 500$  km s$^{-1}$, whose origin should be ascribed to other mechanisms, as discussed below.

\subsection{\bf Stacking the spectra}\label{sec:spectra}

\begin{figure*}
        \centering
        \includegraphics[width=1\textwidth]{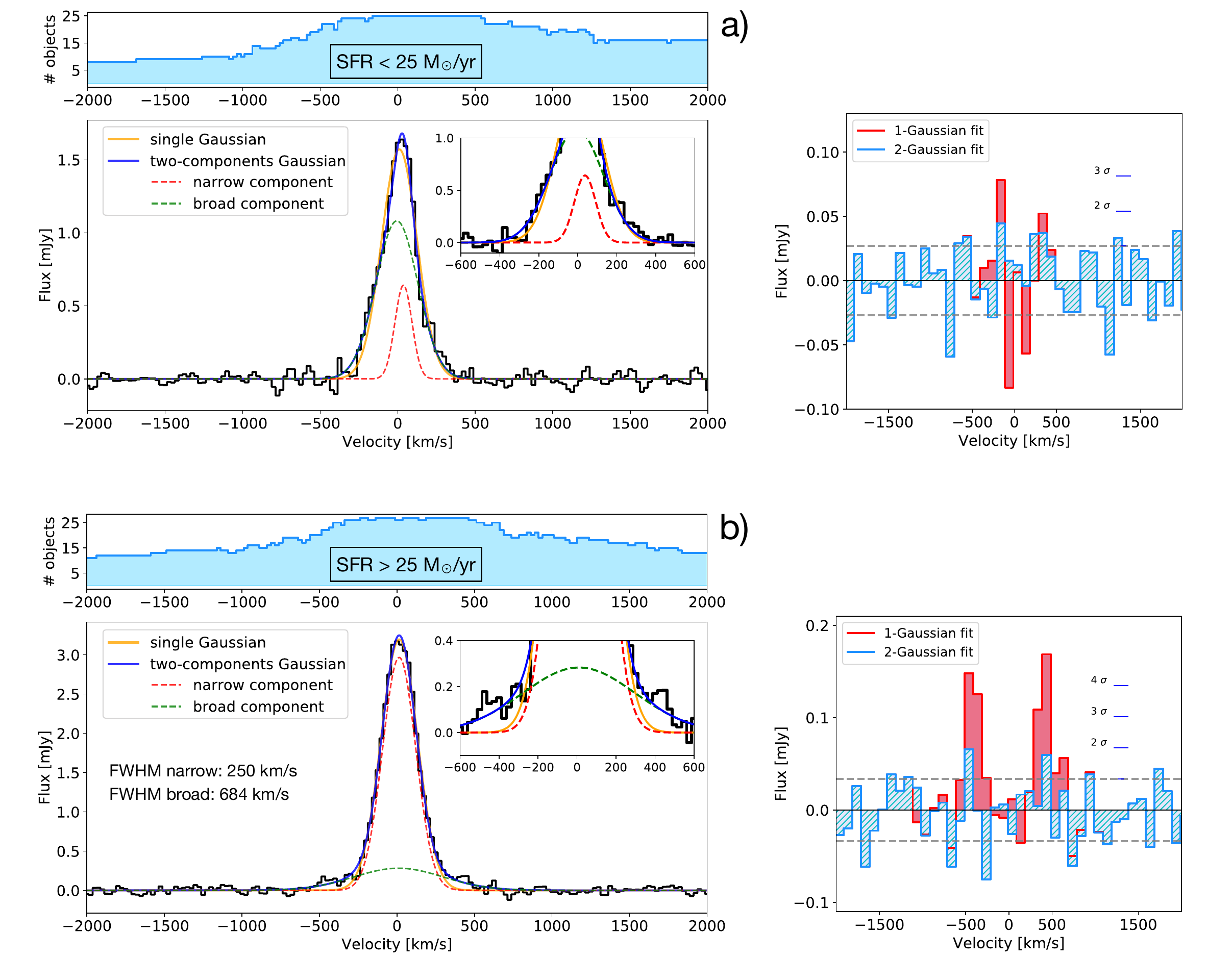}
        \caption{Same description as in Fig. \ref{fig:spectrum_1}. In this case the stacked [C II] spectrum and corresponding residuals from a single- and two-Gaussian best-fits are shown for the low-SFR (a), and the high-SFR (b) subsamples, individually.}
        \label{fig:spectrum_2}
\end{figure*}

In Sect. \ref{sec:residuals} we discussed that a single-Gaussian component is not sufficient to correctly model (on average) the [C II] spectra of a representative (see $M_{\star}$ and SFR distributions in Fig. \ref{fig:physical_properties}) population of high-$z$ normal galaxies (Fig. \ref{fig:residuals_1}).
In particular, we found that the deviation from a single-Gaussian model is related to the SFR, with high(low)-significance flux excess found in the stacked residuals of high(low)-SFR galaxies (Fig. \ref{fig:residuals_2}).
%
%
Interestingly, in line with previous similar works (e.g., \citealp{Gallerani2018}), most of the positive signal revealed in the stacked residuals of highly star-forming galaxies arises from almost symmetric high-velocity tails  
(see Fig. \ref{fig:residuals_2}.b), specifically at velocities consistent with those observed through UV spectroscopy in the outflowing gas accelerated by stellar feedback at similar redshifts (e.g., \citealp{Sugahara2019}; see a discussion in Sect. \ref{sec:introduction}).
%
%
This suggests that the observed flux excess can be ascribed to SFR-driven outflows.
However, to corroborate this hypothesis, we need to test whether a two-component Gaussian  model, that is, a combination of a narrow and a broad component (with the latter tracing the outflowing gas; see Sect. \ref{sec:Results}), can better describe our observations.
We therefore performed a variance-weighted stacking of the [C II] spectra of galaxies in our sample to compute (and compare) the residuals of single-Gaussian and two-component Gaussian best fits.
%
In analogy with Eq. \ref{eq:stack_res}, each $i$-th channel of the stacked spectrum $S^{\rm{stack}}_i $ is defined as:
\begin{equation}\label{eq:stack_spec}
        S^{\rm{stack}}_i = \dfrac{\sum_{k=1}^{N} S_{i, k} \cdot w_k}{\sum_{k=1}^{N} w_k} ,
\end{equation}
where $S_{i, k}$  is the [C II] spectrum of the $k$-th galaxy, and the weighting factor $w_k=1/\sigma^2_k$ is calculated as described in Sect. \ref{sec:residuals}.
\newline
\newline
In Fig. \ref{fig:spectrum_1}.a we show the [C II] spectrum resulting from the stacking of our full sample (with a spectral element binning of 25 km s$^{-1}$) along with the single- and two-component Gaussian best fits.
As for the figures in Sect. \ref{sec:residuals}, a histogram reporting the number of objects  per channel contributing to the corresponding flux is shown on the top panel.
The stacked [C II] spectrum appears to be clearly characterized by weak (less than 10\% of the line peak-flux) broad wings at velocities of a few hundred kilometers per second (see inset in Fig. \ref{fig:spectrum_1}.a).
We find that while a single-Gaussian fit produces significant positive residuals at $v  \sim \pm [300 : 500]$ km s$^{-1}$ (in agreement with results from Sect. \ref{sec:residuals}), a two-component Gaussian fit can accurately describe the stacked spectrum, leaving residuals that are reasonably consistent with simple noise (no peaks exist at $>2\sigma$; see Fig. \ref{fig:spectrum_1}.b).
Our  best fit with a two-Gaussian model results in a combination of a narrow component and a relatively less prominent broad component, in agreement with typical line profiles observed in the presence of outflows at low-$z$ or in galaxies hosting an AGN (see a discussion in Sect. \ref{sec:introduction}).
Both the narrow and broad Gaussian components are centered at the stacked [C II] line peak ($v_{\rm cen} \sim 0 \pm 10$ km s$^{-1}$).
We measure a
FWHM  = $233 \pm 15$ km s$^{-1}$ for the narrow component, and a FWHM = $531 \pm 90$ km s$^{-1}$ for the broad component%
\footnote{Here and in the following, the reported FWHM values are deconvolved for the intrinsic spectral resolution of the stacked spectra (25 km s$^{-1}$).},
where the uncertainties are estimated through a bootstrap analysis, as described in the following paragraphs.
\newline
\newline
As done with the stacked residuals, we test the dependence of the average [C II] spectral line shape on the SFR, dividing our sample into two SFR bins as described in Sect. \ref{sec:residuals}, and repeating the analysis in each group.

We find that the stacked [C II] spectrum of low-SFR galaxies (SFR $< 25$ ${\rm M_{\odot}}$ yr$^{-1}$) does not show clear signs of broad wings (see Fig. \ref{fig:spectrum_2}.a). 
        As expected, given the different line shapes of the constituent spectra (which combine as a sum of Gaussians with different widths) and the larger number of free parameters, the residuals left by the two-component Gaussian best fit are lower than in the single-Gaussian case.
        However, the residuals produced by the single-Gaussian best fit are generally consistent with the noise (no peaks at $\gtrsim3\sigma$), indicating that the  stacked spectrum of the low-SFR galaxies can be sufficiently well described by a single-component Gaussian profile.
        Moreover, in this case the two-component Gaussian best fit is not determined by the expected combination of a narrow and (less prominent) broad component, and therefore does not provide a meaningful result;

The stacked [C II] spectrum of high-SFR galaxies (SFR $> 25$ ${\rm M_{\odot}}$ yr$^{-1}$) shows clear signs of broad wings on the high-velocity tails, at $v~\pm \sim500$ km s$^{-1}$ (see Fig. \ref{fig:spectrum_2}.b).
        The single-Gaussian best fit leaves significant residuals (with peaks exceeding $4\sigma$) in  the velocity range $v  \sim \pm [300 : 500]$ km s$^{-1}$, more prominent than the residuals found in the combined spectrum from the full sample.
        On the other hand, the two-component Gaussian best fit, resulting in the combination of a narrow (FWHM = $251 \pm 10$ km s$^{-1}$) and a broad (FWHM = $684 \pm 75$ km s$^{-1}$) component (see Table 1), produces residuals that are fully consistent with the noise.
\newline

\begin{figure*}
	\centering
	\includegraphics[width=1\textwidth]{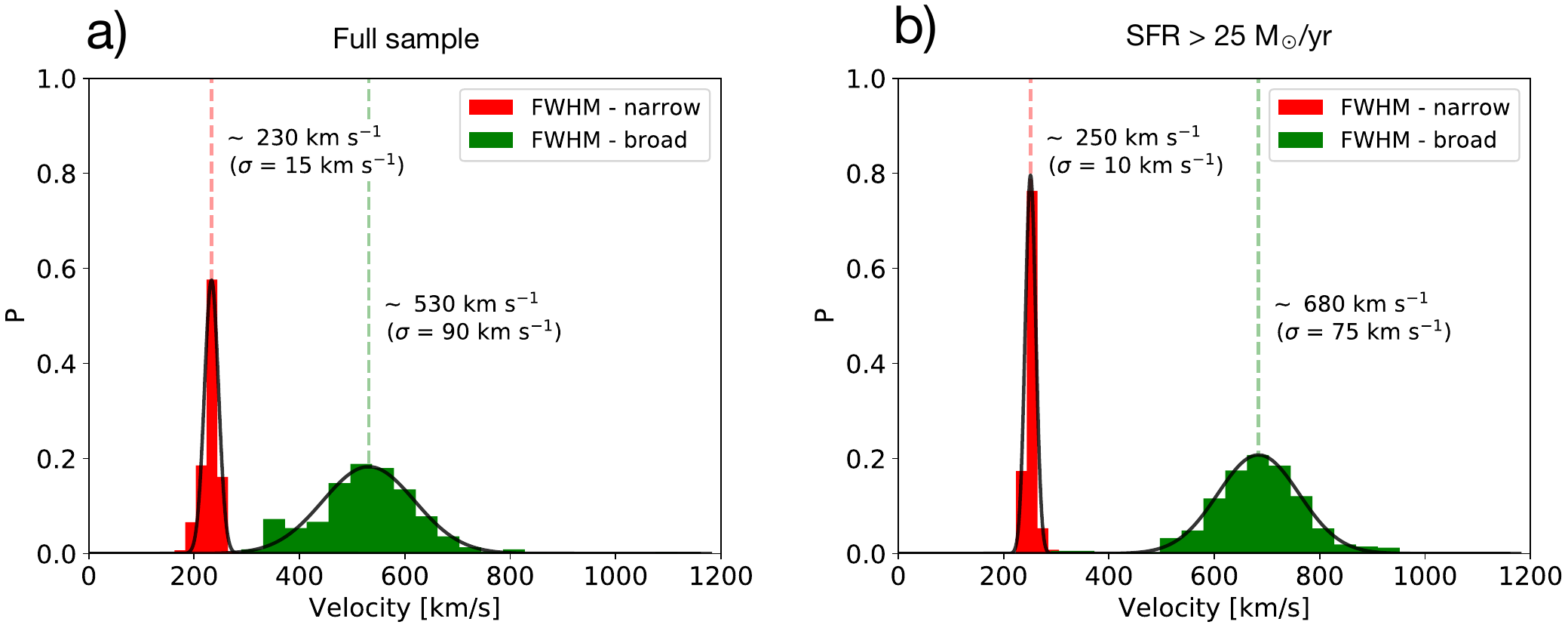}
	\caption{Distribution of the FWHM of narrow (red histograms) and broad (green histograms) components obtained with a bootstrap analysis for the full sample (a) and the high-SFR group (b). The dashed lines represent the median values of the distribution.}
	\label{fig:spectrum_3}
\end{figure*}

\begin{table}\label{tab:table}
        \centering
        \begin{tabular}{lll}
                \hline
                ~  & ~& ~ \\ [-1.5ex]
                ~ & Full Sample & High-SFR group \\  [+0.5ex]
                \hline 
                ~  & ~& ~ \\ [-1.5ex]
                SFR$_{\rm med}$ & 25 ${\rm M_{\odot}}$ yr$^{-1}$ & 50 ${\rm M_{\odot}}$ yr$^{-1}$ \\ [+0.5ex]
                FHWM - narrow & $233 \pm 15$ km s$^{-1}$ & $251 \pm 10$ km s$^{-1}$ \\ [+0.5ex]
                FHWM - broad & $531 \pm 90$ km s$^{-1}$  & $684 \pm 75$ km s$^{-1}$ \\ [+1ex]
                \hline
                ~  & ~& ~ \\
        \end{tabular}
        \caption{Median SFRs of full and high-SFR (sub)samples are summarized, along with the FWHMs of both narrow and broad Gaussian components.
        The uncertainties on the FWHMs are estimated through a bootstrap analysis (see text).}
\end{table}

These findings do not prove the absence of a broad component (i.e., a possible signature of outflows) in the stacked spectrum of the low-SFR subsample. 
Indeed, since [C II] is generally fainter in low-SFR galaxies (see e.g., \citealp{Capak2015, Carniani2018, Matthee2019}; Schaerer et al., in prep.), we might expect this feature to be less evident and most likely below the detection limit. 
On the other hand, since the noise level in both stacks is comparable (given the same number of galaxies in the two bins), we can safely argue that high-SFR galaxies are (on average) characterized by larger and more prominent broad components in their [C II] spectra.
\newline 
\newline
We performed a {bootstrap} analysis to estimate the possible effect of contamination from a few individual sources in the (sub)samples where we found evidence of broad wings.
In Figs. \ref{fig:spectrum_3}.a and \ref{fig:spectrum_3}.b, we show the FWHM distributions of both narrow and broad components obtained with the bootstrap technique for the full sample and the subsample of highly star-forming galaxies, respectively. 
The two panels show that for both narrow and broad components, the peaks of FWHM distributions are highly consistent with the values reported in the analysis described above (without any random replacement), indicating that no obvious dominance by a small number of galaxies is affecting our results.
To improve the reliability of our measurements, we adopt the $\sigma$ of the bootstrapped FWHM distributions (see labels in Fig. \ref{fig:spectrum_3}) as uncertainties on the FWHM values of the stacked spectra reported above (see also Table 1).  
\newline
\newline
We also repeated the stacking of high-SFR galaxies by combining the [C II] spectra extracted from a fixed 4 $\times$ 4 pixel-sized aperture (diameter of $0.6 ''$, i.e., slightly more than half of the averaged circularized beam) centered on the brightest pixels of the velocity-integrated [C II] maps. 
Apart from a small difference in terms of absolute signal, we do not find clear deviations from the stacked spectrum shown in Fig. \ref{fig:spectrum_2}.b, further suggesting that a possible contamination from faint satellites (at least on scales comparable with the beam) does not contribute significantly in building-up the observed broad component.
\newline
\newline
While the FWHM of the broad component that we measure (FWHM $<700$ km s$^{-1}$; see Table 1) is much smaller than that observed in typical high-$z$ QSOs (i.e., FWHM $\gtrsim 2000$ km s$^{-1}$; see \citealp{Maiolino2012,Cicone2015, Bischetti2018}),   
we note that    
some contribution to the [C II] broadening 
may come from 
(i) winds powered by gas accretion onto moderately massive black holes or, (ii) strongly obscured AGNs, especially in the high-SFR group.
One of the objectives of our survey will indeed be to characterize the AGN activity of ALPINE galaxies 
using, for instance, (i) X-ray stacking diagnostics and (ii) stacked UV spectra to constrain the Type II-AGN-sensitive lines (e.g., HeII-$\lambda 1640$ \AA~ and  CIII]-$\lambda 1908$ \AA; see e.g., \citealp{Nakajima2018, LeFevre2019}).
In addition to this, JWST will help in terms of resolved BPT diagram classification (\citealp{Baldwin1981}) and observations of broad H$\alpha$ or [OIII]-$\lambda 5007$ \AA~ line emissions.

\subsection{\bf Stacking the cubes}\label{sec:cubes}

\begin{figure*}
        \centering
        \includegraphics[width=0.8\textwidth]{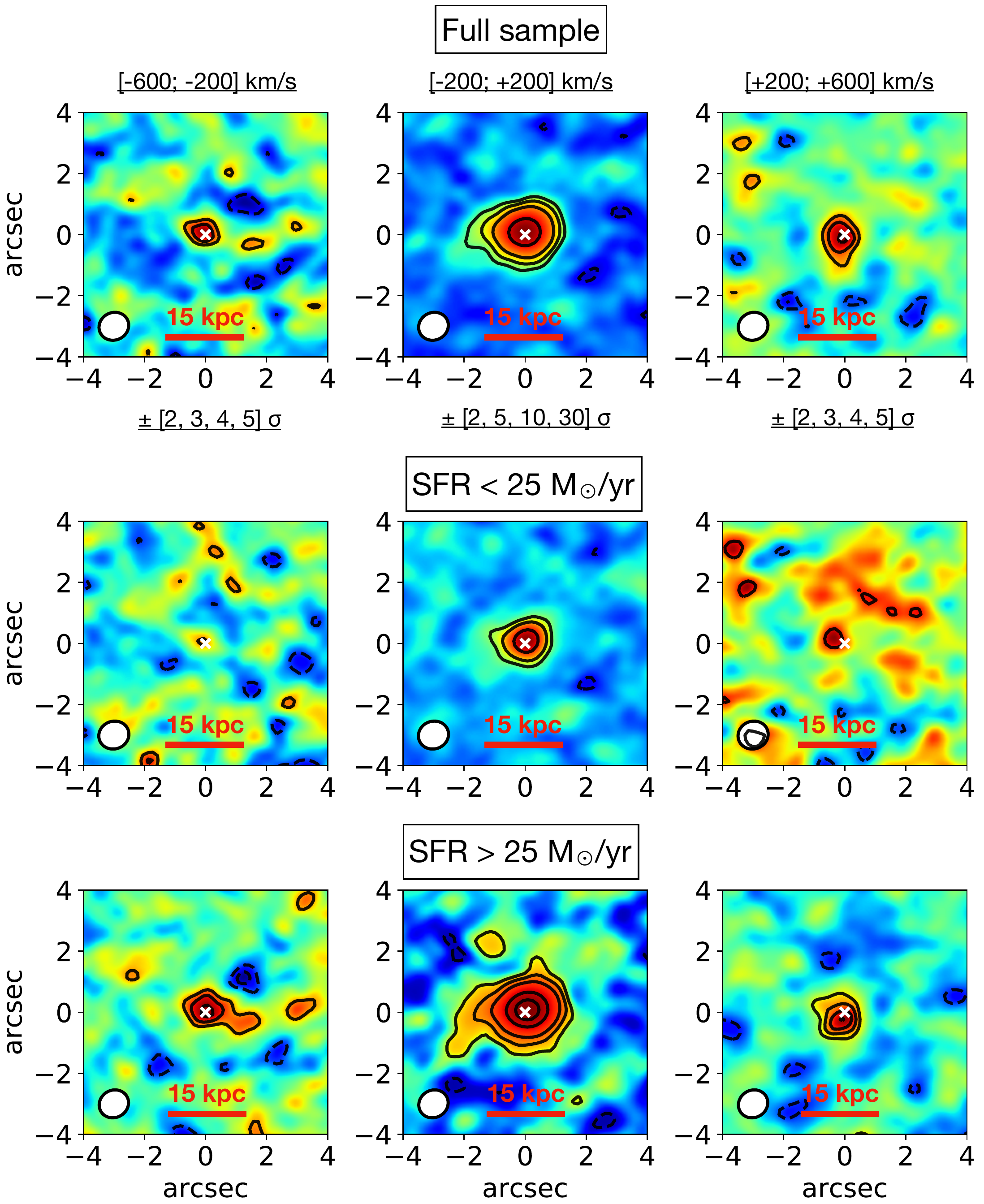}
        \caption{Velocity-integrated [C II] flux maps (central $8'' \times 8''$ regions) are shown in different velocity ranges, from the combined cubes obtained by stacking (a) the full sample, (b) the low-SFR, and (c) the high-SFR groups.
                Left and right panels are representative of the high-velocity tails of the [C II] emission ($[-600: -200]$ and $[+200: +600]$ km s$^{-1}$, respectively), while maps in the central panels trace the [C II] core ($[-200: +200]$  km s$^{-1}$).
                Significance levels of the black contours are reported below the panels of (a). The average synthesized beam is shown in the lower-left corners, while a reference size-scale of 15 kpc is reported at the bottom of each panel. 
                The center of each panel is marked with a white cross.
        }
        \label{fig:cubes_1}
\end{figure*}

As discussed in the previous sections, our stacking analysis of [C II] spectra shows that the significance of 
residuals from a single-Gaussian fit
and broad wings on the high-velocity tails, at $v~\pm \sim500$ km s$^{-1}$, 
increases with the SFR, indicating that star-formation-driven outflows are at play in high-$z$ normal galaxies.
To better characterize the outflow properties, we explored the morphologies and spatial extensions of both the core and high-velocity wings of the [C II] line. 
We combined the [C II] cubes of our galaxies $C_i$, spectrally and spatially aligned as discussed in Sect. \ref{sec:extraction}, following a vector variance-weighted stacking: 
\begin{equation}\label{eq:stack_cube}
        C^{\rm{stack}}_i = \dfrac{\sum_{k=1}^{N} C_{i, k} \cdot w_{i, k}}{\sum_{k=1}^{N} w_{i, k}} .
\end{equation}
Equation \ref{eq:stack_cube} is a generalized version of Eq. \ref{eq:stack_spec} (see e.g., \citealp{Fruchter2002, Bischetti2018}), where $C^{\rm{stack}}_i$ is the stacked cube composed by $i$ slices,  $C_{i, k}$ is the [C II] cube of the $k$-th galaxy and $w_{i, k}$ is the weighting factor defined as $w_{i, k}=1/\sigma^2_{i, k}$. 
Here $\sigma^2_{i, k}$ is defined as the spatial rms estimated from a large emission-free region at each $i$-th slice of each $k$-th galaxy, and allows us to account for any frequency-dependent noise variation in the [C II] cubes.
\newline
\newline
Following the same procedure described in Sects. \ref{sec:residuals} and  \ref{sec:spectra}, we performed the stacking analysis of [C II] cubes for both (i) the full sample and (ii) two groups of galaxies with SFR higher or lower than the median SFR in our sample, that is SFR $\lessgtr 25$ ${\rm M_{\odot}}$ yr$^{-1}$.
We then collapsed the spectral slices of the [C II] stacked cubes 
in the velocity ranges (i) $[-200: +200]$  km s$^{-1}$, and (ii) $[-600: -200]$,  $[+200: +600]$ km s$^{-1}$.
Those ranges were specifically chosen to produce velocity-integrated flux maps of (i) the core of the [C II] emission, and (ii) the [C II] high-velocity tails, respectively.
In Fig. \ref{fig:cubes_1} we show the central $8'' \times 8''$ regions of the flux maps from the stacked cubes.

We find that [C II] emission is detected up to 4$\sigma$ in the velocity-integrated maps at $[-600: -200]$ and $[+200: +600]$ km s$^{-1}$ of the full sample, and up to 5$\sigma$ in the high-SFR group. 
        Only tentative detections ($\sim 2\sigma$) are revealed in the high-velocity tails of the low-SFR group (see side panels of Fig. \ref{fig:cubes_1}).
        Where detected, the high-velocity [C II] emission is marginally resolved (compared with the average beam of the observations in the stack%
        \footnote{The stacked synthesized beam of our observations has a major axis FWHM of $0.98''$, a minor axis FWHM of $0.89'',$ and a position angle of $-30$ deg.}%
        ), extending on beam-deconvolved%
        \footnote{We calculated beam-deconvolved sizes by fitting a 2D Gaussian model and subtracting in quadrature the major and minor axis FWHM of our stacked synthesized beam.} %
        angular sizes of $\sim0.9''$, corresponding to $\sim 6$ kpc at $z_{\rm med}= 5$ (the median redshift of our galaxies);

In the velocity-integrated image at $[-200: +200]$  km s$^{-1}$, which traces the core of line, we detect [C II] emission at exceptionally high significance in all our three (sub)samples, 
        that is, $\gtrsim30\sigma$ in the full sample and high-SFR bin and $\gtrsim10\sigma$ in the low-SFR group (see central panels of Fig. \ref{fig:cubes_1}).
        Interestingly, while in all three cases [C II] emission is fully resolved and extended on angular scales of $> 2''$ ($> 15$ kpc at $z_{\rm med}= 5$), the core of [C II] line emission appears to be more extended for the high-SFR galaxies, with low-S/N ($2\sigma$) features extending up to angular scales of $\gtrsim 3''$, corresponding to about $20$ kpc at $z_{\rm med}=5$.

To test the reliability of these results, we repeated the analysis carrying out a {median} stacking instead of the variance-weighted mean stacking described in Eq. \ref{eq:stack_cube}. 
We do not find evident deviations, confirming that our findings are not affected by outliers in the distribution.
\newline
\begin{figure}
        \centering
        \includegraphics[width=1\columnwidth]{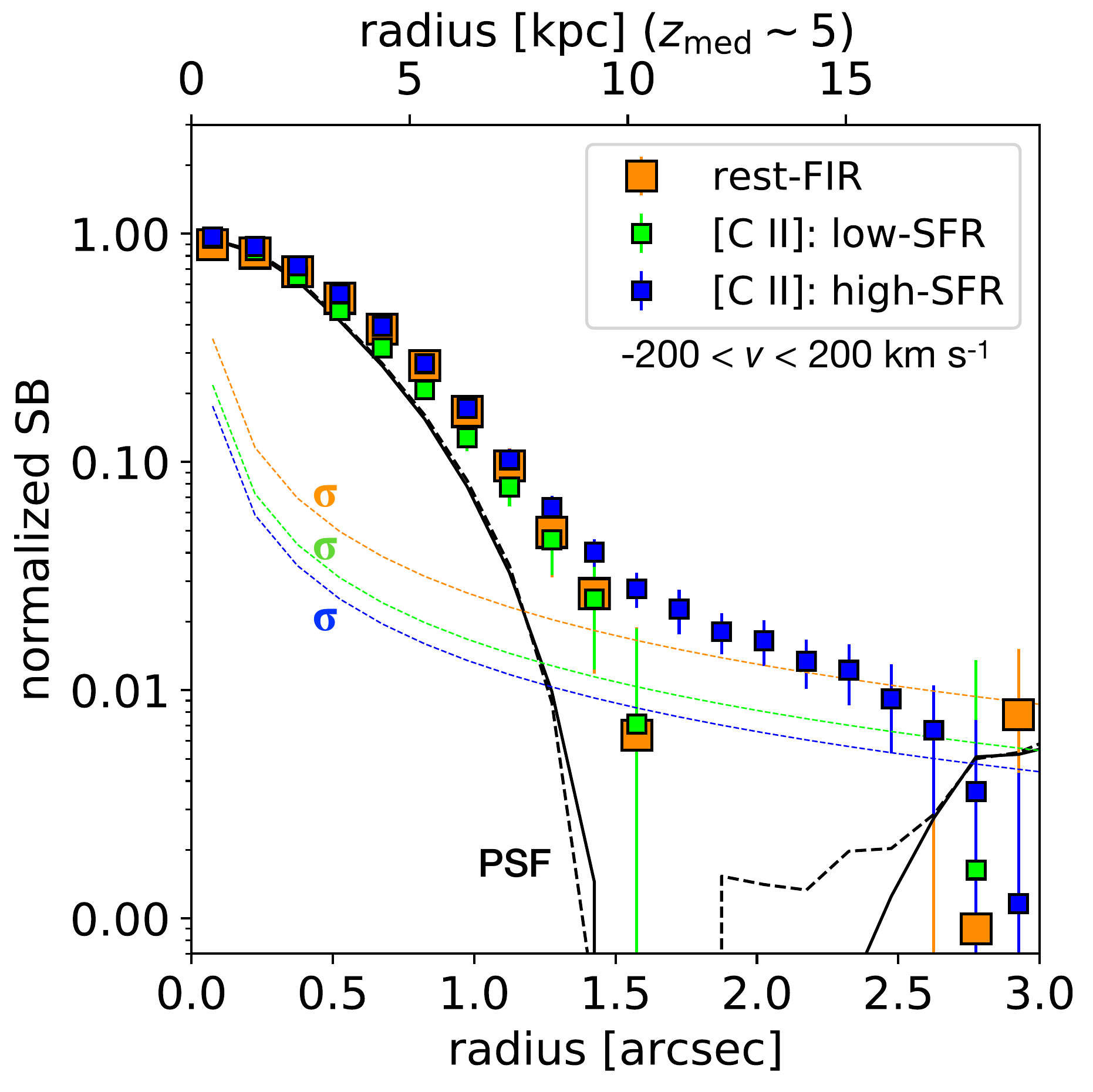}
        \caption{Circularly averaged radial profiles computed in concentric 0.3$''$-binned annuli are shown for (i) the stacked PSF of our ALMA observations (black solid line for galaxies in our sample, and black dashed line for ALPINE continuum-detected galaxies), (ii) the stacked FIR continuum of 23 ALPINE continuum-detected galaxies (mostly belonging to the high-SFR group, see text; orange squares), and (iii) the stacked maps of [C II] cores ($[-200: +200]$  km s$^{-1}$) for galaxies in the low-SFR (green squares) and high-SFR (blue square) groups. Error bars are indicative of the $\pm~1\sigma$ dispersion of fluxes in each annulus, while the thin dashed lines represent the Poissonian noise associated with the radial profiles\protect\footnotemark. %
        }
        \label{fig:cubes_2}
\end{figure}
\footnotetext{We estimated the Poissonian noise level by dividing the rms of the normalized [C II]-flux (or continuum) images  by the square root of each annulus area.}

To constrain the typical extension of the stacked [C II] line core with higher accuracy and quantify its dependence on the SFR (as suggested by the flux maps in Fig. \ref{fig:cubes_1}), we computed the circularly averaged radial profiles of surface brightness (SB) from the low-velocity [C II] flux maps of our stacked (sub)samples (see Fig. \ref{fig:cubes_2}).
We then compared them with the radial profiles of SB extracted from the stacked point spread function (PSF) image and the stacked FIR continuum;
the former is obtained by stacking the ALMA PSF cubes of galaxies in our sample (using Eq. \ref{eq:stack_cube}) and by collapsing the channels at $[-200: +200]$  km s$^{-1}$,
while the latter is obtained through a mean- and rms-weighted stack of the FIR-continuum images of the 23 ALPINE continuum-detected galaxies ($\sim 90\%$ of which belong to our high-SFR subsample; see details in B\'{e}thermin et al. in prep., and Khusanova et al., in prep.).
Figure \ref{fig:cubes_2} shows that
the radial profile of the stacked [C II] line core in the low-SFR group is slightly more extended than the average PSF radial profile. 
It extends similarly to the FIR continuum (deconvolved effective radii of $\sim1.2''$; $\sim 8$ kpc at $z_{\rm med}=5$), suggesting that they are both tracing gas emitted on the same (galactic) scale.

On the other hand the radial profile of the stacked  [C II] line core in the high-SFR subsample can be seen to extend well beyond the analogous emission from lower-SFR galaxies. The stacked FIR continuum, reaching a deconvolved effective radius of $2.3''$, corresponds to a physical distance of $\sim 15$ kpc at $z_{\rm med}=5$.
        While the relatively more compact profile of low-SFR galaxies could in principle be interpreted as an effect of limited sensitivity,        
        we can safely argue that higher-SFR galaxies show larger [C II] fluxes on average on radial scales $>10$ kpc.
\newline

Our statistical detection of a low-velocity [C II] emission extended on such large physical sizes (diameter scales of $\sim 30$ kpc) suggests 
the existence of metal enriched circumgalactic halos surrounding main sequence high-$z$ galaxies, confirming with larger statistics and significance the result obtained by \cite{Fujimoto2019}, who found a 20 kpc (diameter scale) [C II] halo in the stacked cube of 18 galaxies at $5<z<7$ (see their discussion for an overview of the theoretical mechanisms proposed to explain the extended emission).
%
%
Since outflows of processed material are needed to enrich with carbon the primordial CGM of early systems (see \citealp{Fujimoto2019}), 
the detected [C II] halo is evidence of (i) past star-formation-driven outflows and (ii) gas mixing at play in the CGM of high-$z$ normal star-forming galaxies (see a discussion in Sect. \ref{sec:discussion}).
%
We postpone further explorations of these findings to future papers, including, for example, analyses of the rest-frame UV continuum (Fujimoto et al., in prep) and Ly$\alpha$ stacked emissions, as well as comparisons with tailored hydrodynamical simulations (e.g., \citealp{Behrens2019, Pallottini2019}; Mayer et al., in prep.).

\section{Discussion}\label{sec:discussion}

The stacking analysis of [C II] spectra and cubes of ALPINE galaxies described in Sect. \ref{sec:Results} suggests that 
outflows are unequivocal in normal star-forming galaxies at $4<z<6$ (see possible caveats in Sect. \ref{sec:contaminants}). 
Interestingly, we find that the intensity and the significance of [C II] emission in the broad wings at the high-velocity tails of the stacked spectra and cubes increases with the SFR, 
supporting the star-formation-induced nature of the observed outflows.

\subsubsection*{\it Mass outflow rate and efficiency of star-formation-driven outflows}

To estimate the efficiency of star-formation-driven outflows at play in high-$z$ galaxies, we calculated the 
mass outflow rate ($\dot{M}_{\rm out}$) following an approach similar to previous studies of outflows in the [C II] spectra of QSOs and normal galaxies (e.g., \citealp{Maiolino2012, Cicone2015, Janssen2016, Gallerani2018,Bischetti2018}).

We use the luminosity of the broad [C II] component to get an estimate of the outflowing atomic gas mass, $M_{\rm out}^{\rm atom}$, adopting the relation from \cite{Hailey2010}:
\begin{equation} \label{eq:m_outfl}
        \begin{aligned}
                \dfrac{M_{\rm out}^{\rm atom}}{\rm M_\odot} = 0.77 \left(       \dfrac{0.7 ~ L_{\rm[C II]}}{\rm L_\odot} \right)  \times \left(    \dfrac{1.4\times10^{-4}}{X_{\rm C^+}}\right) \times
                \\
                \dfrac{1+2 e^{-91~{\rm K}/T} +  n_{\rm crit}/n}{2 e^{-91~{\rm K}/T} },
        \end{aligned}
\end{equation}
where $X_{\rm C^+}$ is the abundance of ${\rm C^+}$ per hydrogen atom, $n$ is the gas number density, $n_{\rm crit}$ is the critical density of the [C II] 158 $\mu$m transition (i.e., $\sim 3 \times 10^3$ cm$^{-3}$), and $T$ is the gas temperature.
This relation is derived under the assumption that most of the broad [C II] emission arises from atomic gas (see a discussion in \citealp{Janssen2016}); 
        specifically, 
        70\% of the total [C II] flux (corresponding to the factor 0.7 in the first parenthesis of Eq. \ref{eq:m_outfl}) arises from photodissociation regions (PDRs; e.g., \citealp{Stacey1991, Stacey2010}), with only the remaining fraction arising from other ISM phases (see e.g.,  \citealp{Cormier2012, Vallini2015, Vallini2017, Lagache2018, Ferrara2019}, for discussions on the relative contribution of various gas phases). It is also assumed that the [C II] emission is optically thin; this sets a lower limit on $M_{\rm out}^{\rm atom}$ since, in case of optically thick [C II], the actual outflowing gas mass would be larger.
%
We use Eq. \ref{eq:m_outfl} assuming 
(i) a gas number density higher than $n_{\rm crit}$ (this approximation gives a lower limit on the mass of the atomic gas, as discussed in \citealp{Maiolino2012}),
and 
(ii) a ${\rm C^+}$ abundance, $X_{\rm C^+}\sim1.4\times10^{-4}$, (\citealp{Savage1996}) and 
a gas temperature in the range %
%
$T\sim60 - 200~{\rm K}$, both typical of PDRs (see e.g., \citealp{Kaufman1999, Hollenbach1999, Wolfire2003, Kaufman2006}).
%
%
Applying Eq. \ref{eq:m_outfl}  to the stacked [C II] spectra of our full and high-SFR (sub)samples (where broad components are detected) we infer a 
mass of the outflowing atomic gas, 
$M_{\rm out}^{\rm atom} = (2.1 \pm 0.8) \times 10^8 ~{\rm M_{\odot}}$, for the full sample, and
$M_{\rm out}^{\rm atom} = (2.9 \pm 1.2) \times 10^8 ~{\rm M_{\odot}}$ for the high-SFR subsample,
%
%
where the errors reflect the explored range of $T$  and the uncertainties on $L_{\rm[C II]}$.
We then compute the atomic $\dot{M}_{\rm out}$ assuming time-averaged expelled shells or clumps (\citealp{Rupke2005, Gallerani2018}):
\begin{equation}
        \dot{M}_{\rm out} = \dfrac{v_{\rm out} ~ M_{\rm out}}{R_{\rm out}},
\end{equation}
where:
\newline
- $v_{\rm out}$ is the typical velocity of the atomic outflowing gas traced by [C II]. 
We adopt $v_{\rm out} \sim 500$ km s$^{-1}$, based on the velocity scale at which we observe significant peaks of deviation from a single-Gaussian model in the stacked residuals and spectra (see Fig. \ref{fig:residuals_1}, \ref{fig:residuals_2}.b and \ref{fig:spectrum_1}.b);
\newline
- $R_{\rm out}$ is the typical spatial extension of the outflow-emitting regions. 
We use as an estimate $R_{\rm out} \sim 6$ kpc, adopting the beam-deconvolved sizes derived in Sect. \ref{sec:cubes} from the high-velocity [C II] emission in the stacked cubes of the full and high-SFR (sub)samples (high-velocity [C II] emission is only tentatively detected in the low-SFR bin; see Fig. \ref{fig:cubes_1}).
Furthermore, we estimate mass outflow rates of $\dot{M}_{\rm out} = 18\pm5$  ${\rm M_{\odot}}$ yr$^{-1}$  for the full sample, and $\dot{M}_{\rm out} = 25\pm8$   ${\rm M_{\odot}}$ yr$^{-1}$  for the high-SFR subsample.
These values are lower than the median SFRs measured in the two bins, namely 
SFR$_{\rm med} = 25$ ${\rm M_{\odot}}$ yr$^{-1}$ and SFR$_{\rm med} = 50 $ ${\rm M_{\odot}}$ yr$^{-1}$ 
in the full sample and the high-SFR group, respectively.
%
However, we emphasize that our estimate only accounts for the atomic gas phase of the outflow, while a significant fraction of the outflowing gas is likely to be in the molecular and ionized form, as commonly observed in local star forming galaxies (e.g., \citealp{Veilleux2005,Heckman2017,Rupke2018}).
For example, recent work by \cite{Fluetsch2019}, who study multi-phase outflows in a sample of local galaxies and AGNs, shows that when including all the gas phases, the total mass-loss rate increases roughly by up to 0.5 dex with respect to the value estimated from the atomic outflow only, suggesting that a coarse estimation of the total $\dot{M}_{\rm out}^{\rm tot}$ can be obtained multiplying the  $\dot{M}_{\rm out}$ measured in the atomic phase by a factor of three.
%
%
%
%

%
Assuming that similar considerations apply to our sample of high-$z$ normal galaxies, we estimate total mass outflow rates of $\dot{M}_{\rm out}^{\rm tot} \sim 55\pm15$ ${\rm M_{\odot}}$ yr$^{-1}$ for the full sample, and  $\dot{M}_{\rm out}^{\rm tot} \sim 75\pm24$ ${\rm M_{\odot}}$ yr$^{-1}$  for the high-SFR group.
%
\newline \newline
\begin{figure}
        \centering
        \includegraphics[width=1\columnwidth]{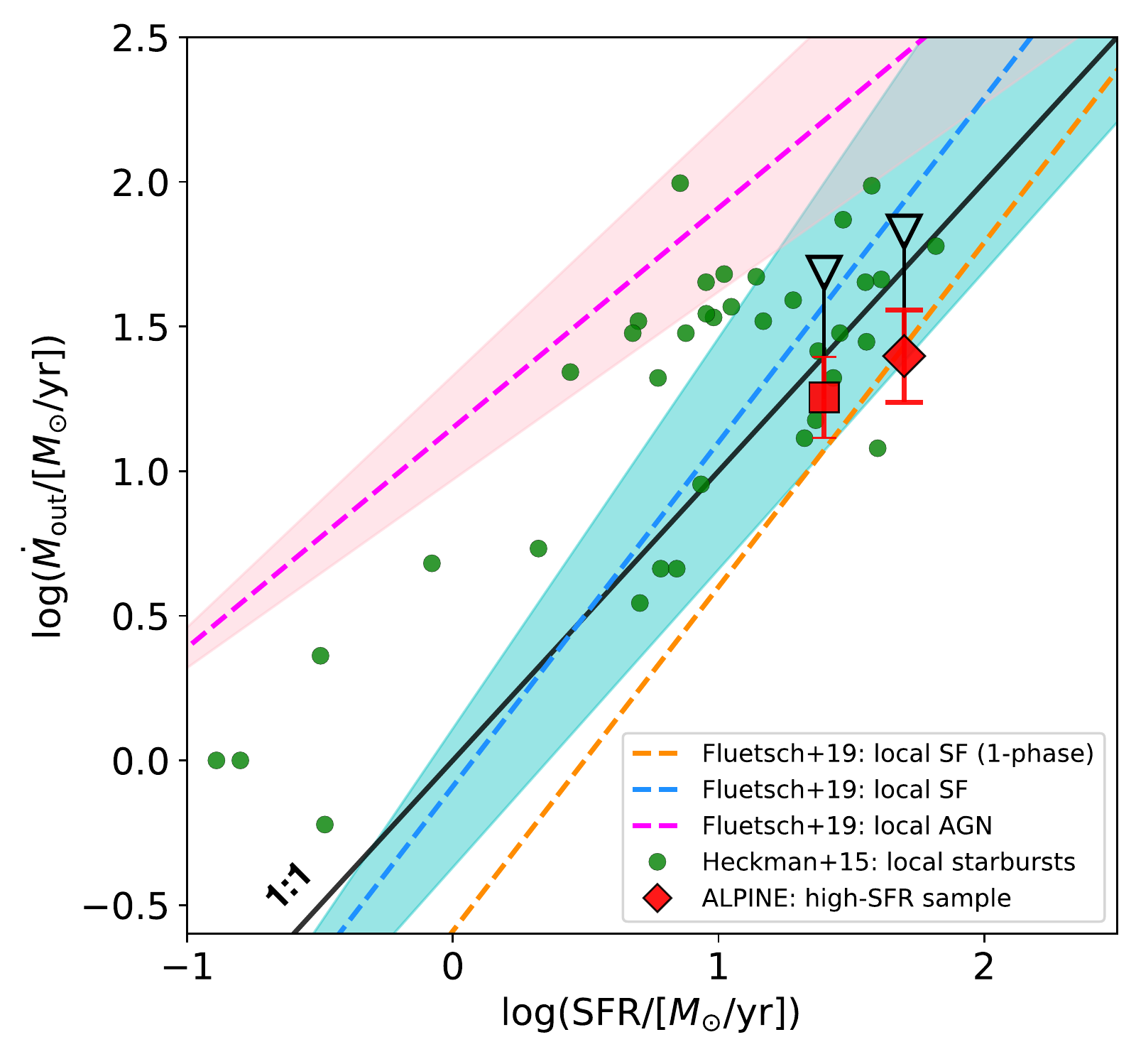}
        \caption{Comparison of our results with a compilation of data at low $z$ in the log($\dot{M}_{\rm out}$)-log(SFR) diagram.
                The red square (diamond) indicates the average outflow rate for the atomic component obtained from the stacking of the full (high-SFR) sample, while  red bars indicate the associated uncertainty ($\pm1\sigma$).
                The black arrows show our estimate of the total $\dot{M}_{\rm out}$ (calculated adopting a correction for multi-phase outflows; see text).
                The orange (cyan) dashed line indicates the best fits for single-phase (three-phases, i.e., molecular, ionised, and atomic) $\dot{M}_{\rm out}$ observations in local star-forming galaxies, while the magenta dashed line represents the best fit from observation of local AGNs (from \citealp{Fluetsch2019}).
                Filled colored regions indicate the $2\sigma$ dispersion around the best fits.
                The green points show the distribution of a sample of local starbursts (\citealp{Heckman2015}).
                The blue solid line indicates the 1:1 relation ($\eta = 1$).
        }
        \label{fig:discussion}
\end{figure}
In Fig. \ref{fig:discussion} we show a comparison of our results with a compilation of local starbursts (\citealp{Heckman2015}) and the best-fitting relations of local AGNs and normal star-forming galaxies from \cite{Fluetsch2019} in the log($\dot{M}_{\rm out}$)--log(SFR) diagram.
We find that 
our [C II] observations 
yield mass-loading factors, $\eta= \dfrac{\dot{M}_{\rm out}}{\rm SFR}$, lower than (or consistent with) unity ($\eta^{\rm atom}\sim0.3 - 0.9$), in analogy with what is found in local star-forming galaxies (see e.g., \citealp{GarciaBurillo2015,Cicone2016,Fluetsch2019, RodriguezDelPino2019}; see the orange line).
Assuming corrections for the multi-phase outflowing gas contribution (using the calibration discussed above; see blue dashed line in Fig. \ref{fig:discussion}) we find higher mass-loading factors (see black arrows) in the range $\eta^{\rm tot}\sim1 - 3$, still below the $\eta$ observed in local AGNs ($\eta^{\rm AGN}>5$;
see e.g., \citealp{Fluetsch2019,Fiore2017} for a discussion on the dependence of  $\eta^{\rm AGN}$ on AGN properties).
Therefore, even assuming that all the gas phases significantly contribute to the outflowing gas, the total mass-loss rate produced by star-formation-driven outflows still remains roughly comparable with the SFR.

This suggests that star-formation feedback, when compared to AGN feedback, is a less efficient mechanism for rapid {quenching} of normal galaxies in the early Universe.
While possibly contributing to the regulation of star formation as in local galaxies (see e.g., \citealp{Montero2019}), it is likely not the dominant factor needed to explain the observed population of passive galaxies at $z\sim2-3$ (e.g., \citealp{Merlin2018, Santini2019, Valentino2019}).

\subsubsection*{\it Intergalactic and circumgalactic metal enrichment}

%
It is still not clear whether or not star formation-driven outflows can actually escape the DM halos and therefore effectively remove the fuel for future star formation.
On the one hand, the sensitivity of currently available data (even in the deepest integrations with ALMA, tracing both atomic and molecular FIR lines) is far from being sufficient at revealing the spatial extension of the star-formation-driven winds around individual high-$z$ main sequence galaxies on scales comparable with their virial radii.
On the other hand, while the stacking of ALPINE-like large samples can provide significantly improved sensitivity, the randomness of wind directions and geometries strongly challenges the detection of spatially extended outflowing gas (as seen in Sect. \ref{sec:cubes}, where the  high-velocity [C II] flux is more compact than the core component).
\newline
\newline
Another way to figure out the fate of the outflows is to compare their typical velocities, $v_{\rm outf}$, with the escape velocities, $v_{\rm esc}$, of the DM halos.
We estimate $v_{\rm esc}$ of the DM halos hosting the galaxies in our sample, using the formula:
\begin{equation}\label{eq:v_esc}
        v_{\rm esc} = \sqrt{\dfrac{2~G~M_{\rm DM}}{r_{\rm DM}}},
\end{equation}
where $r_{\rm DM}$ is the virial radius and $M_{\rm DM}$ is the mass of the halo.
We calculate $r_{\rm DM}$ using the commonly adopted hypothesis of virialized halos (see e.g., \citealp{Huang2017}):
\begin{equation}\label{eq:r_dm}
        r_{\rm DM} = \left[             \dfrac{3 ~ M_{\rm DM}}{4 ~\pi ~200~\rho_{\rm crit}(z)}               \right]^{1/3} , 
\end{equation}
where $\rho_{\rm crit}(z)$ is the critical density of the Universe at redshift $z$ 
and $M_{\rm DM}$ was estimated using empirically calibrated stellar mass-halo mass (SMHM) relations (see e.g.,  \citealp{Behroozi2013,Durkalec2015,Behroozi2019}).
%
%
%
%
%
%

Galaxies in the high-SFR group of our sample, where (as discussed in Sect. \ref{sec:Results}) the signatures of atomic star-formation-driven outflows are  unequivocal, have stellar masses in the range $M_{\star} =10^{10}-10^{11.2}  ~{\rm M_{\odot}}$. 
Those stellar masses, according to the SMHM relation by \cite{Behroozi2019}, correspond to DM halos masses in the range $M_{\rm DM} \sim 7\times 10^{11} - 5 \times 10^{12} ~ {\rm M_{\odot}}$ and virial radii of $r_{\rm DM}\sim40-100$ kpc (Eq. \ref{eq:r_dm}
).
Therefore, using Eq. \ref{eq:v_esc}, we find typical escapes velocities of $v_{\rm esc}\sim 400 - 800$ km s$^{-1}$. 
These values of $v_{\rm esc}$, compared with the outflow velocities ($v_{\rm out}\lesssim500$ km s$^{-1}$) found in our stacked [C II] spectrum, 
suggest that 
a fraction of gas accelerated by star-formation-driven outflows may escape the halo only in less massive galaxies (and possibly contribute to the IGM enrichment, as expected by models; e.g., \citealp{Oppenheimer2010, Pallottini2014, Muratov2015}), while
this is unlikely to happen for the more massive galaxies.
The outflowing gas that cannot escape the halo would instead be \textit{trapped} in the CGM and eventually virialize after mixing with both the quiescent and the inflowing primordial gas, producing the large reservoir of enriched circumgalactic gas that we observe in [C II] on scales of $\sim 30$ kpc (see Sect. \ref{sec:cubes}; see also a discussion in \citealp{Fujimoto2019}).
Altogether these results confirm the expectations of cosmological simulations (see e.g., \citealp{Somerville2015,Hopkins2014,Hayward2017}) that the baryon cycle and the enriched gas exchanges with the CGM are at work in normal galaxies already in the early Universe. 

\section{Conclusions}\label{sec:conclusions}

Here we present the stacking analysis of the [C II] emission detected by ALMA in 50 main sequence star-forming galaxies at $4 < z < 6$ (see information on the sample in Sect. \ref{sec:Observations} and Sect. \ref{sec:contaminants}) drawn from the ALPINE survey (\citealp{LeFevre2019a}; B\'{e}thermin et al. 2019; Faisst et al. 2019).
The combination of (i) a large statistical sample and (ii) a wealth of ancillary multi-wavelength photometry (from UV to FIR) provided by ALPINE sets the ideal conditions to progress in studying the efficiency of star-formation-driven feedback and circumgalactic enrichment at early epochs.
Our main findings can be summarized as follows.

\begin{itemize}
        
\item   
To check whether the [C II] line profiles of our galaxies can be sufficiently well described by a single-Gaussian model, we performed a variance-weighted stacking analysis of the  [C II] residuals, computed by subtracting a single-component Gaussian fit to each [C II] spectrum (see Sect. \ref{sec:residuals}). 
We observe typical deviations from a single-component Gaussian model, consisting of flux excesses (with peaks at $>4\sigma$) in the high-velocity tails of the stacked residuals, at $|v| \lesssim 500$ km s$^{-1}$ (Fig. \ref{fig:residuals_1} and Fig. \ref{fig:residuals_2}.b), in line with previous similar studies carried out on smaller samples (see \citealp{Gallerani2018}).
\newline 
\item
We performed a variance-weighted stacking of the [C II] spectra (see Sect. \ref{sec:spectra})
and find that the stacked [C II] profile of normal star-forming galaxies in our sample is characterized by typical signatures of outflows in its high-velocity tails.
In particular, we detect broad wings at velocities of a few hundred kilometers per second (Fig. \ref{fig:spectrum_1}), and find that the average [C II] spectrum can be accurately described by a two-component Gaussian fit (analogous to observations of QSOs; e.g.,  \citealp{Maiolino2012,Cicone2015,Bischetti2018}), resulting in a combination of a narrow component (FWHM $\sim 230$ km s$^{-1}$) and a relatively less prominent broad component (FWHM $\sim 530$ km s$^{-1}$).
\newline
\item 
We repeated the [C II] residuals and spectra stacking dividing our sample into two equally populated SFR-defined bins, using SFR$_{\rm med}$ = 25 ${\rm M_{\odot}}$ yr$^{-1}$ as a threshold.
We find that both (i) the significance of deviation from a single-component Gaussian model in the combined residuals (Fig. \ref{fig:residuals_2}) and (ii) the significance of the broad wings in the high-velocity tails of the stacked [C II] spectrum (Fig. \ref{fig:spectrum_2}) increase (decrease) when stacking the subsample of high (low)-SFR galaxies, confirming the star-formation-driven nature of these features. 
In particular, the stacked [C II] spectrum of high-SFR galaxies shows a broad component with a FWHM of $\sim 700$ km s$^{-1}$.
\newline
\item 
We constrain the efficiency of star-formation-driven outflows at early epochs estimating the resulting mass-outflow rates (see Sect. \ref{sec:discussion}). 
We find values roughly comparable with the SFRs ($\dot{M}_{\rm out} \lesssim 30$   ${\rm M_{\odot}}$ yr$^{-1}$), yielding a mass loading factor lower than (or consistent with) unity ($\eta^{\rm atom}\lesssim1$), similarly to what is found in local, normal star-forming galaxies (Fig. \ref{fig:discussion}; see e.g., \citealp{Cicone2016, Fluetsch2019}). 
Even when considering a contribution to the outflow from multiple gas phases, the estimated mass loading factor ($\eta^{\rm tot}\sim1 - 3$) is still below the $\eta$ observed in AGNs, suggesting that stellar feedback may play a lesser role in quenching galaxies at $z>4$ and producing passive galaxies by $z\sim2-3$.
\newline
\item 
To better characterize the outflow properties and explore morphologies and spatial extensions of both the core and the high-velocity wings of the [C II] emission, we performed a stacking analysis of the datacubes (see Sect. \ref{sec:cubes}).
We find that the combined [C II] core emission ($|v| < 200$  km s$^{-1}$) of galaxies in the high-SFR subsample extends over physical sizes of $\sim$ 30 kpc (diameter scale), well beyond the stacked FIR continuum and the [C II] core emission of lower-SFR galaxies (Fig. \ref{fig:cubes_2}).
\newline
The detection of such extended metal-enriched gas, likely tracing circumgalactic gas enriched by past outflows, corroborates previous, similar studies (see \citealp{Fujimoto2019}; see also Fujimoto et al., in prep.), confirming that the baryon cycle, metal circulation, and gas mixing in the CGM are at work in normal star-forming galaxies in the early Universe.

\end{itemize}


\section*{Acknowledgements}
The authors would like to thank the anonymous referee for the useful suggestions.
M.G. would like to thank Andrea Ferrara, Simona Gallerani, Andrea Pallottini, Stefano Carniani, Jorryt Matthee, Emanuele Daddi, Andreas Schruba and Roberto Decarli for helpful discussions.
G.C.J. and R.M. acknowledge ERC Advanced Grant 695671 "QUENCH'' and support by the Science and Technology Facilities Council (STFC).
M.B. acknowledges FONDECYT regular grant 1170618.
E.I.\ acknowledges partial support from FONDECYT through grant N$^\circ$\,1171710.
F.L., C.G., F.P. and M.T. acknowledge the support from  a grant PRIN MIUR 2017.
L.V. acknowledges funding from the European Union’s Horizon 2020 research and innovation program under the Marie Skłodowska-Curie Grant agreement No. 746119. 
S.T. acknowledges support from the ERC Consolidator Grant funding scheme (project Con-TExT, grant No. 648179). 
The Cosmic Dawn Center is funded by the Danish National Research Foundation under grant No. 140.
R.C. acknowledges financial support from CONICYT Doctorado Nacional N$^\circ$\,21161487 and CONICYT PIA ACT172033.
J.D.S is supported by the JSPS KAKENHI Grant Number *JP18H04346*, and the World Premier International Research Center Initiative (WPI Initiative), MEXT, Japan.
This paper is based on data obtained with the ALMA Observatory, under Large Program 2017.1.00428.L. 
ALMA is a partnership of ESO (representing its member states), NSF (USA) and NINS (Japan), together with NRC (Canada), MOST and ASIAA (Taiwan), and KASI (Republic of Korea), in cooperation with the Republic of Chile. 
The Joint ALMA Observatory is operated by ESO, AUI/NRAO and NAOJ. 
This program receives financial support from the French CNRS-INSU Programme National Cosmologie et Galaxies.

\bibliographystyle{aa}
\bibliography{biblio.bib}

\end{document}